\documentclass[aps, prb, reprint]{revtex4-1} %{revtex4-1}%
\usepackage{mathtools}
\usepackage{amssymb}  % include amsfonts package
\usepackage{amsfonts}  % American Math Society fonts, define \mathbb, \mathfrak
\usepackage{amsmath}  % multi-lines formulars, \cfrac
\usepackage{amsthm}  % provide a PROOF enviroment
\usepackage{bm}  % bold math
\usepackage{bbm}  % hollow math
\usepackage[]{hyperref}
\usepackage{graphicx}
\usepackage{tabularx}
\usepackage{bbding}
\usepackage{float}
\usepackage{comment}

\theoremstyle{plain}

\usepackage{tikz, datatool, ifthen, csvsimple}
\usetikzlibrary{arrows, shapes, math, decorations, decorations.markings}
\usetikzlibrary{calc}
\usepackage{pgfplots}
\pgfplotsset{compat=newest}
\pgfplotsset{
    colormap={mycm}{rgb255=(225, 225, 225) rgb255=(225, 225, 225)},
    colormap/mycm/.style={
        colormap name=mycm,
    },
}
\usepackage{pgfmath}
\input pgfmath.tex

\newcommand{\imth}{\hspace{1pt}\mathrm{i}\hspace{1pt}}
\newcommand{\bea}{\begin{eqnarray}}
\newcommand{\eea}{\end{eqnarray}}
\newcommand{\phs}{\tilde{\mathcal{C}}}
\newcommand{\bpm}{\begin{pmatrix}}
\newcommand{\epm}{\end{pmatrix}}
\newcommand{\calI}{\mathcal{I}}

\begin{document}

\title{Unconventional quantum phase transitions\\ in a one-dimensional Lieb-Schultz-Mattis system}
\author{Wayne Zheng$^{1, 3}$, D. N. Sheng$^2$, and Yuan-Ming Lu$^1$}
\affiliation{$^1$Department of Physics, The Ohio State University, Columbus, Ohio 43210, USA \\
$^2$Department of Physics and Astronomy, California State University Northridge, Northridge, California 91330, USA \\
$^3$Department of Physics, The Chinese University of Hong Kong, Shatin, New Territories, Hong Kong, China}

%%\author{Yuan-Ming Lu}
%%\affiliation{Department of Physics, The Ohio State University, Columbus, Ohio 43210, USA}

\date{\today}

\begin{abstract}
We study quantum phases and phase transitions in a one-dimensional interacting fermion system with a Lieb-Schultz-Mattis (LSM) type anomaly.
Specifically, the inversion symmetry enforces any symmetry-preserving gapped ground state of the system to be a Kitaev chain, following a Lieb-Schultz-Mattis type theorem that we prove.
Alternatively, via the Jordan-Wigner transformation, this system describes a spin system whose gapped ground states must break either the inversion or the Ising symmetry associated with fermion parity. We obtain a phase diagram using analytical methods and variational matrix product state simulations, and study the critical behaviors of the quantum phase transitions therein using entanglement entropy, energy variance and finite size scaling of order parameters. In particular, we observe continuous phase transitions between different ordered phases that are beyond the Ginzburg-Landau-Wilson paradigm, in analogy to the deconfined quantum critical points in two spatial dimensions.
We show this type of 1D deconfined quantum critical point is described by the Tomonaga-Luttinger liquid theory, and extract the Luttinger parameter and critical exponents.
We also identify a gapless phase between two ordered phases, which cannot be described by a U(1) Luttinger liquid.
\end{abstract}

\maketitle
\tableofcontents

\section{Introduction}

A paradigm beyond the Landau theory of spontaneous symmetry breaking is the \emph{deconfined quantum critical point} (DQCP). It was firstly proposed for the Neel order to valence bond solid (VBS) transition on a two-dimensional square lattice \cite{PhysRevB.70.144407, Senthil1490}, as a type of continuous quantum phase transition between two ordered phases that cannot be related by symmetry breakings. Compared to the Ginzburg-Landau-Wilson paradigm, a DQCP features many novel aspects such as emergent symmetries and self-duality~\cite{PhysRevX.7.031051}.

Recently a lot of interests arise for revisiting one spatial dimension (1d) to realize the deconfined quantum criticality. In particular, the 1d spin-$1/2$ chain with both nearest- and second-neighbor anisotropic exchange interactions have been extensively studied\cite{PhysRevB.99.205153,PhysRevB.99.075103, PhysRevB.99.165143, PhysRevB.100.125137}, which exhibits a DQCP between a (anti-)ferromagnetic order and a VBS phase. The critical behaviors of this DQCP has also been carefully examined and compared to field-theory predictions. 
The phase transition between these two gapped orders is a direct second-order quantum phase transition, whose long-wavelength low-energy theory is expected to exhibit an emergent $U(1)$ symmetry.
On the other hand, this 1d DQCP is closely related to the Lieb-Schultz-Mattis (LSM) theorem~\cite{LIEB1961407,PhysRevLett.84.1535,PhysRevX.6.041068, PhysRevLett.119.127202,  LU2020168060, PhysRevB.69.104431, PhysRevB.98.125120, 1705.04691, Parameswaran2013}, which forbids a gapped symmetric ground state that preserves both translation and the discrete $Z_2\times Z_2$ spin rotational symmetries~\cite{PhysRevB.99.075103}.

In this paper, we study a 1D lattice model of interacting fermions, with a different LSM-type anomaly. In particular, any gapped ground state that preserves a site-centered inversion symmetry must be a Kitaev chain, with an odd number of Majorana bound states on each boundary. Through a Jordan-Wigner transformation, it becomes a spin-$1/2$ chain, whose gapped ground states must break either the inversion symmetry or the Ising symmetry associated with the fermion parity. We prove such a LSM-type theorem, and study a generic 1d fermion model with nearest-neighbor couplings that preserves this inversion symmetry. The phase diagram of our model has a rich structure: there are DQCPs between different ordered phases beyond the Landau theory, as well as stable gapless phases separating the ordered phases. The rest part of this paper is organized as follows: In Sec.~\ref{sec:model}, we show our model and discuss its symmetries.
In Sec.~\ref{sec:numerical_methods}, the numerical methods used to study the model are discussed.
In Sec.~\ref{sec:phase_diagram}, the phase diagram of the model is obtained, using analytical solutions in the non-interacting limit and numerical results for the interacting model.
In Sec.~\ref{sec:critical_behaivor}, the critical behaviors at the phase boundaries are carefully analyzed, focusing on the DQCP described by the Luttinger liquid theory, and a stable gapless phase. Finally, the concluding remarks are given in Sec.~\ref{sec:conlusion}.

\section{The model \label{sec:model}}

\subsection{Lattice model and its symmetries}

We consider the following 1D model of interacting fermions
\begin{equation}
    \begin{aligned}
        H
        &=\sum_{j}(-)^{j}\left[t\left(c_{j}^{\dagger}c_{j+1}+h.c.\right)+\left(\Delta c_{j}^{\dagger}c_{j+1}^{\dagger}+h.c.\right)\right] \\
        &+\left(\text{i}\Delta^{\prime}c_{j}^{\dagger}c_{j+1}^{\dagger}+h.c.\right)+V\left(n_{j}-\frac{1}{2}\right)\left(n_{j+1}-\frac{1}{2}\right)
    \end{aligned}
    \label{eq:fermion_ham}
\end{equation}
It breaks all the global (onsite) symmetries except for the fermion parity conservation $\mathcal{P}_f=(-1)^{\hat F}$. In fact, assuming $t>0$, this Hamiltonian includes all possible nearest-neighbor (NN) couplings that preserve a site-centered unitary inversion symmetry defined as follows:
\begin{equation}\label{eq:inversion}
  c_j\overset{\mathcal{I}}\longrightarrow\imth c_{-j}^\dagger.
\end{equation}
If $c_{j}\equiv\gamma_{j}+\imth\eta_{j}$ is written in terms of Majorana fermions $\gamma_{j}$ and $\eta_{j}$, this very inversion symmetry actually realize
\begin{equation}
    \begin{pmatrix}
        \gamma_{j} \\
        \eta_{j}
    \end{pmatrix}
    \overset{\mathcal{I}}\longrightarrow
    \begin{pmatrix}
        \eta_{-j} \\
        \gamma_{-j}
    \end{pmatrix}
\end{equation}
saying permuting the two corresponding Majorana fermions.
In a periodic chain with even sites, the above NN-only model also preserves a magnetic translation symmetry $\tilde T_x$ defined as
\bea\label{eq:magnetic translation}
&\tilde T_x\equiv(-1)^{\sum_jj\hat n_j}T_x\cdot\mathcal{K},~~~n_j\equiv c_j^\dagger c_j;\\
\notag &c_j\overset{\tilde T_x}\longrightarrow(-1)^{j+1}c_{j+1}.
\eea
where $\mathcal{K}$ represents the complex conjugation. This gives rise to a link-centered anti-unitary inversion symmetry $\tilde{\mathcal{I}}=\mathcal{I}\cdot\tilde T_x$:
\bea
c_j\overset{\tilde{\mathcal{I}}}\longrightarrow(-1)^{j+1}\imth c^\dagger_{-j-1}
\eea
Besides, the model Eq.~(\ref{eq:fermion_ham}) also exhibits an anti-unitary particle-hole symmetry
\begin{equation}\label{eq:particle-hole sym}
\tilde{\mathcal{C}}=\mathcal{C}\cdot\mathcal{K}:~~~c_j\rightarrow (-1)^jc_j^\dagger
\end{equation}
where $\mathcal{C}$ represents the unitary particle-hole transformation.

The generators $\{\mathcal{I},\tilde{\mathcal{I}}\}$ of the symmetry group satisfy the following algebra:
\bea\label{eq:kramers inv}
\tilde{\mathcal{I}}^2=1,\\
\label{eq:inv+fermion parity}\mathcal{I}(-1)^{\hat F}\mathcal{I}^{-1}=(-1)^L(-1)^{\hat F}.
\eea
where $L\in{\mathbb{Z}}$ is the system size of the 1d chain. In a periodic chain of length $L=0\mod2$, the magnetic translation satisfies:
\bea
(\tilde T_x)^L=(\mathcal{P}_f)^{L(L-1)/2}
\eea

Using the following Jordan-Wigner transformation
\bea
\notag&c_j=(-)^{\frac{j(j-1)}2}\left[\prod_{k<j}\left(-\sigma_{k}^{z}\right)\right]\sigma_j^-,\\
&S_j^z\equiv\frac{\sigma^{z}_j}2=c_j^\dagger c_j-\frac12\label{eq:jordan-wigner}
\eea
the fermion model Eq.~(\ref{eq:fermion_ham}) can also be rewritten as a spin-$\frac12$ chain:
\begin{equation}
\begin{aligned}
&H_\text{spin}
=\sum_{j}\left(t \sigma_{j}^{+}\sigma_{j+1}^{-}+\Delta \sigma_{j}^{+}\sigma_{j+1}^{+}+h.c.\right)\\
&+(-)^{j}\left(\text{i}\Delta^{\prime}\sigma_{j}^{+}\sigma_{j+1}^{+}+h.c.\right)+VS_j^zS^z_{j+1}\\
&=\sum_j\sum_{\alpha=x,y,z}J_\alpha S_j^\alpha S_{j+1}^\alpha+(-1)^j\Gamma(S_j^xS_{j+1}^y+S_j^yS^x_{j+1})
\end{aligned}
\label{eq:spin_ham}
\end{equation}
where the exchange couplings are given by
\bea
&\notag J_x=2(t+\Delta),~~J_y=2(t-\Delta),~~J_z=V,\\
&\Gamma=-2\Delta^\prime.
\eea
This is a familiar XYZ model\cite{PhysRevB.99.205153}, supplemented by an extra staggered anisotropic exchange coupling of strength $\Gamma$. The introduction of $\Gamma$ terms has important consequences: it leads to new gapless phases unseen in the XYZ model\cite{PhysRevB.99.205153,PhysRevB.99.075103}.

The symmetry group of the spin model\footnote{Strictly speaking, the symmetry operators (\ref{sym:spin:inversion}) in the spin model differs from the inversion symmetry (\ref{eq:inversion}) in the fermion model, since the latter is not a locality-preserving unitary. This subtlety is discussed in detail in Appendix \ref{app:symmetry}. } are generated by
\bea\label{sym:spin:ising}
&\mathcal{P}_f=(-1)^{\hat F}=\prod_r(-\sigma^{z}_r),\\
&\tilde{T_x}=T_x\cdot\mathcal{K},\\
&\phs=(\prod_r\sigma^{x}_r)\cdot\mathcal{K},\\
&\mathcal{I}=(\prod_r \sigma^{x}_r)\cdot\mathcal{O}_I.\label{sym:spin:inversion}
\eea
where $\mathcal{O}_I$ is the spatial inversion operator, and the Pauli matrix $\sigma^{x}_r=2 S_r^x$. In particular, the inversion symmetry $\mathcal{I}$ anticommutes with the Ising symmetry Eq.~(\ref{sym:spin:ising}), on a spin chain of an odd length.

\subsection{A Lieb-Schultz-Mattis theorem for the Kitaev chain}

One significant consequence of the inversion symmetry Eq.~(\ref{eq:inversion}) is the following theorem of Lieb-Schultz-Mattis~\cite{LIEB1961407, Hastings_2005, PhysRevLett.84.1535, Parameswaran2013, PhysRevLett.114.077201, PhysRevB.101.224437} (LSM) type:\\

{\bf Theorem:} \emph{In an one-dimensional spinless fermion system preserving the inversion symmetry (\ref{eq:inversion}), any gapped symmetric ground state must be a Kitaev chain.} \\

This theorem is closely related to the family of LSM theorems for symmetry protected topological (SPT) phases~\cite{PhysRevB.87.155114, annurev-conmatphys-031214-014740} discussed recently~\cite{PhysRevB.98.125120, LU2020168060, 1705.04691, 1907.08596}, but differs in the sense that here any gapped symmetric ground state is enforced to be a nontrivial invertible phase~\cite{1604.06527}, i.e. the Kitaev chain~\cite{Kitaev_2001}, rather than SPT phases.

Below we prove the theorem in two aspects. First we show that when restricted to a non-interacting fermion system with translation symmetry, we can use the polarization formula of BdG bands to show a gapped ground state must have a nontrivial $\mathbb{Z}_2$ topological invariant, hence belonging to a Kitaev chain. Next, we will show that in a generic interacting open chain with inversion symmetry, a gapped ground state must be a Kitaev chain with Majorana zero modes at the boundary.

Firstly, we consider a periodic fermion chain with an even number of sites. Since each unit cell includes two sites, they transform as
\bea
\bpm c_{2r}\\c_{2r+1}\epm\overset{\mathcal{I}}\longrightarrow\imth\bpm c^\dagger_{-2r}\\c^\dagger_{-2(r+1)+1}\epm
\eea
In the basis of Majorana fermions
\bea
c_{2r}=\frac{\chi_{r}+\imth\eta_r}2,~~c_{2r+1}=\frac{\chi^\prime_r+\imth\eta^\prime_r}2.
\eea
they transform as follows under inversion
\bea
\Phi_k\equiv\bpm\chi(k)\\ \eta(k)\\ \chi^\prime(k)\\ \eta^\prime(k)\epm\overset{\mathcal{I}}\longrightarrow\bpm0&1&0&0\\1&0&0&0\\0&0&0&e^{\imth k}\\0&0&e^{\imth k}&0\epm\Phi_{-k}
\eea
in the momentum space. In other words, the inversion symmetry is implemented by unitary rotation
\bea
R_\mathcal{I}(k)=\bpm1&0\\0&e^{\imth k}\epm_{\vec\tau}\otimes\mu_x
\eea
where $\vec\tau$ and $\vec\mu$ are Pauli matrices for the sublattice and Majorana indices respectively. A generic quadratic Bogoliubov-de Gennes (BdG) Hamiltonian has the following form
\bea
\hat H_\text{free}=\sum_{0\leq k\leq\pi}\psi_{-k}^T h(k)\psi_k,
\eea
where Hermitian matrix $h(k)$ satisfies particle-hole and inversion symmetries:
\bea\label{eq:bdg:phs}
&h^T(k)=h^\ast(k)=-h(-k),\\
&R_\mathcal{I}(k)h(k)R_\mathcal{I}^{-1}(k)=h(-k).\label{eq:bdg:inversion}
\eea
The $\mathbb{Z}_2$-valued topological invariant~\cite{PhysRevB.78.195125, PhysRevB.88.075419} for such a BdG Hamiltonian in symmetry class D is given by the quantized polarization of the filled BdG bands:
\bea
&\nu=e^{\imth\int\text{d}kA(k)}=\pm1,\\
&\notag A(k)=\imth\sum_{\epsilon_n(k)<0}\langle \psi_{n}(k)|\partial_k\psi_n(k)\rangle.
\eea
where $A(k)$ is the Berry connection of the filled bands. The quantized $\nu=\pm1$ is a consequence of the particle-hole symmetry (\ref{eq:bdg:phs}), and inversion symmetry (\ref{eq:bdg:inversion}) constrains the Berry curvature as follows:
\bea
\notag&A(-k)=-\imth\sum_{\epsilon<0}\langle \psi_{n}(k)|R_\mathcal{I}^\dagger(k)\partial_kR_\mathcal{I}(k)|\psi_{n}(k)\rangle-A(k)\\
&=\sum_{\epsilon<0}\langle \psi_{n}(k)|\frac{1+\tau_z}2|\psi_{n}(k)\rangle-A(k)
\eea
Particle-hole symmetry (\ref{eq:bdg:phs}) further indicates that
\bea\notag
\sum_{\epsilon<0}\langle \psi_{n}(k)|\frac{1+\tau_z}2|\psi_{n}(k)\rangle=\sum_{\epsilon>0}\langle \psi_{n}(-k)|\frac{1+\tau_z}2|\psi_{n}(-k)\rangle
\eea
and hence
\bea
\notag&\int\text{d}kA(k)=\pi\text{Tr}(1+\tau_z)-\int\text{d}kA(k)\Longrightarrow\\
&\int\text{d}kA(k)=\pi\mod2\pi\\
&\Longrightarrow\nu=e^{\imth\int\text{d}kA(k)}=-1.
\eea
Therefore, for a gapped BdG Hamiltonian with a well-defined polarization, the $\mathbb{Z}_2$ invariant must be nontrivial, and hence it must be a Kitaev chain with Majorana edge modes.

Secondly, we consider a generic interacting Hamiltonian preserving inversion symmetry Eq.~ (\ref{eq:inversion}), on an open chain with an odd number of sites $L=1\mod 2$. In this case, there is one single inversion center on the middle site, and the inversion symmetry acts as a supersymmetry which changes the fermion parity:
\bea\label{eq:susy}
\calI(-1)^{\hat F}\calI^{-1}=-(-1)^{\hat F}
\eea
This implies at least 2-fold degeneracy for all energy levels, hence two degenerate ground states on an open chain with opposite fermion parities. If the bulk is gapped, this necessarily leads to zero modes on the edge. Since there is no extra global symmetry in the system to protect the edge modes, they can only be an odd number of Majorana zero modes (MZMs) on each edge. This indicates the ground state is an open Kitaev chain. And its total fermion parity is flipped by the inversion symmetry, which exchanges the MZMs on the two edges.

Therefore, we have shown that a gapped ground state preserving inversion symmetry Eq.~ (\ref{eq:inversion}) must be a Kitaev chain with an odd number of MZMs on each open boundary. This LSM theorem for Kitaev chain in the fermion context can also be translated into the spin chain language, via Jordan-Wigner transformation Eq.~(\ref{eq:jordan-wigner}). In the spin language, it manifests as the more familiar LSM theorem:

{\bf Theorem:} \emph{In an one-dimensional spin-$1/2$ chain with both Ising symmetry Eq.~(\ref{sym:spin:ising}) and inversion symmetry Eq.~(\ref{sym:spin:inversion}), its ground state is either gapless or spontaneously breaks symmetries.} \\

In other words, the spin-$1/2$ chain with both Ising and inversion symmetries do not admit any short-range entangled ground state.

\section{Numerical methods \label{sec:numerical_methods}}

\subsection{Spin chain representation and order parameters}

For the convenience of numerical simulation, we use the traditional Jordan-Wigner transformation:
\begin{equation}
   \begin{aligned}
       c_{j}&=\left[\prod_{k=0}^{j-1}\left(-\sigma_{k}^{z}\right)\right]\sigma_{j}^{-}, \\
       \sigma_{j}^{z}&=2c_{j}^{\dagger}c_{j}-1,
   \end{aligned}
\end{equation}
to rewrite the fermionic model Eq.~(\ref{eq:fermion_ham}) as a spin chain
\begin{equation}
\begin{aligned}
H
%&=\sum_{j}(-)^{j}\left[\left(t\sigma_{j}^{+}\sigma_{j+1}^{-}+h.c.\right)+\left(\Delta \sigma_{j}^{+}\sigma_{j+1}^{+}+h.c.\right)\right] \\
%&+\left(\text{i}\Delta^{\prime}\sigma_{j}^{+}\sigma_{j+1}^{+}+h.c.\right)+\frac{1}{4}V\sigma_{j}^{z}\sigma_{j+1}^{z} \\
&=\sum_{j}(-)^{j}\left(t\sigma_{j}^{+}\sigma_{j+1}^{-}+h.c.\right) \\
&+\left(\left[(-)^{j}\Delta+\text{i}\Delta^{\prime}\right]\sigma_{j}^{+}\sigma_{j+1}^{+}+h.c.\right) \\
&+\frac{1}{4}V\sigma_{j}^{z}\sigma_{j+1}^{z},
\end{aligned}
\label{eq:spin_ham 1}
\end{equation}
where $\sigma^{\pm}=(\sigma^{x}\pm\text{i}\sigma^{y})/2$.
$\sigma^{x, y, z}$ are Pauli matrices. Note that this Jordan-Wigner transformation differs from Eq. (\ref{eq:jordan-wigner}) by a $(-1)^{j(j-1)}/2$ sign, hence a different form of the Hamiltonian (\ref{eq:spin_ham 1}) compared to (\ref{eq:spin_ham}). In this representation, the symmetry generators of Hamiltonian (\ref{eq:spin_ham 1}) writes:
\bea
\notag&\mathcal{P}_f=(-1)^{\hat F}=\prod_r(-\sigma^{z}_r),\\
\notag&\tilde{T_x}=(\prod_{r=\text{odd}}\sigma^z_r)T_x\cdot\mathcal{K},\\
\notag&\phs=(\prod_r\sigma^{x}_r)\cdot\mathcal{K},\\
\notag&\mathcal{I}=(\prod_r \sigma^{x}_r)\cdot(\prod_{r=\text{odd}}\sigma^z_r)\cdot\mathcal{O}_I.
\eea

The on-site matrix product operator (MPO) for the Hamiltonian Eq.~(\ref{eq:spin_ham}) can be written as a $5\times 5$ matrix $V^{[j]}=$
\begin{equation}
    \begin{pmatrix}
        \mathbbm{1} & 0 & 0 & 0 & 0  \\
        \sigma^{+} & 0 & 0 & 0 & 0  \\
        \sigma^{-} & 0 & 0 & 0 & 0  \\
        \sigma^{z} & 0 & 0 & 0 & 0  \\
        0 & t_{j}\sigma^{-}+\tilde{\Delta}_{j}\sigma^{+} & t_{j}\sigma^{+}+\tilde{\Delta}_{j}^{*}\sigma^{-} & \frac{1}{4}V\sigma^{z} & \mathbbm{1}
    \end{pmatrix},
\end{equation}
where we define $t_{j}\equiv(-)^{j}t, \tilde{\Delta}_{j}\equiv(-)^{j}\Delta+\text{i}\Delta^{\prime}$.
The boundary vectors for the open boundary condition (OBC) are $v_{L}=(0, 0, 0, 0, \mathbbm{1}), v_{R}=(\mathbbm{1}, 0, 0, 0, 0)^{T}$.

Recall that $\mathcal{P}_{f}\rightarrow -\mathcal{P}_{f}$ under the inversion operation $\mathcal{I}$ in an open chain of an odd length, according to (\ref{eq:susy}). As indicated by the LSM theorem earlier, a gapped ground state either breaks inversion $\calI$ or the parity symmetry $\mathcal{P}_f$. If $\mathcal{P}_{f}$ is preserved while $\mathcal{I}$ is broken, it can be characterized by the non-vanishing order parameters such as
\begin{equation}
\begin{aligned}
    M_{\text{FM-}z}
    &\equiv\frac{1}{L}\sum_{j=-N}^{N}\sigma_{j}^{z}, \\
    M_{\text{AFM-}z}
    &\equiv\frac{1}{L}\sum_{j=-N}^{N}(-)^{j}\sigma_{j}^{z},
    \label{eq:op_z}
\end{aligned}
\end{equation}
which are invariant under the symmetry $\mathcal{P}_{f}$ whereas $\mathcal{I}^{-1}M_{\text{(A)FM-}z}\mathcal{I}=-M_{\text{(A)FM-}z}$.
Their MPOs can be written in the form as
\begin{equation}
    V_{\text{FM-}z}^{[j]}=
    \begin{pmatrix}
        \mathbbm{1} & 0 \\
        \sigma^{z} & \mathbbm{1}
    \end{pmatrix}, \quad
    V_{\text{AFM-}z}^{[j]}=
    \begin{pmatrix}
        \mathbbm{1} & 0 \\
        e^{\text{i}\pi{j}}\sigma^{z} & \mathbbm{1}
    \end{pmatrix}.
    \label{eq:}
\end{equation}
To detect possible symmetry breaking of $\mathcal{P}_{f}$, we use the following inversion-symmetric order parameter
\begin{equation}
    M_{x}=\frac{1}{L}\sum_{j=-N}^{N}(-)^{\frac{j(j+1)}{2}}\sigma_{j}^{x}.
    \label{eq:op_x}
\end{equation}
Since $\mathcal{P}_{f}^{-1}M_{x}\mathcal{P}_{f}=-M_{x}$, non-vanishing $\langle{M}_{x}\rangle$ implies the spontaneous symmetry breaking of $\mathcal{P}_{f}$. Although spontaneous symmetry breaking (SSB) cannot really occur on a finite chain, DMRG tends to select a minimally entangled ground state, which means $|\langle{M}\rangle|$ could be a good estimation for the SSB. However, it can be very unstable near a quantum critical point, where a macroscopic superposed cat state becomes possible. To overcome this difficulty, we can add an explicit symmetry-breaking term such as $H_{h}=h\left(\sigma_{-N}^{z}+\sigma_{N}^{z}\right)$ on the boundary to split the degenerate ground states.
Alternatively, we can use root mean square order parameter $\sqrt{\langle{M}^{2}\rangle}$ instead of the order parameter $|\langle{M}\rangle|$ itself to detect SSB~\cite{binder2010, Pang2019}.
In addition to these order parameters, another alternative way to detect spontaneous symmetry breaking is to look at the two point correlation function associated with the corresponding order parameter~\cite{SciPostPhysLectNotes.11}.
The advantage of the correlation function is that it can be used in finite systems where the expectation value of the order parameter is zero.

\subsection{Matrix product state, variance and entanglement}

For Eq.~(\ref{eq:spin_ham}) on a finite chain consisting the number of sites $L$ under OBC, its many-body wavefunction can be represented by a matrix product state (MPS) as
\begin{equation}
    |\psi\rangle
    =\sum_{\left\{s\right\}}\left(A^{s_{0}}A^{s_{1}}\dots{A}^{s_{L-1}}\right)|s_{0}, s_{1}, \dots, s_{L-1}\rangle.
    \label{eq:}
\end{equation}
$A^{s_{0},s_{L-1}}$ are two boundary vectors with dimensions $1\times{d}$ and $d\times{1}$, respectively.
Here $d=2$ denotes the dimension of the local Hilbert space.
The \emph{exact} MPS requires the largest bond dimension $\chi_{\text{max}}=\sqrt{d}^{L}$ at the center of the chain.
However, it is not practically achievable therefore we need to put a fixed cut-off $\chi$ to compress the wavefunction.
Written in the Schmidt basis~\cite{PhysRevLett.101.010504} in terms of two spatial parts $A$ and $B$, the wavefunction looks like
\begin{equation}
    |\psi\rangle
    =\sum_{\alpha=0}^{\chi-1}e^{-\frac{\omega_{\alpha}}{2}}|\alpha\rangle_{A}\otimes|\alpha\rangle_{B},
    \label{eq:schmidt}
\end{equation}
where $\{\omega\}$ is the \emph{entanglement spectrum}, which is the eigenvalue spectrum of the local entanglement Hamiltonian $\mathcal{H}_{A}$ defined by the reduced density matrix $\rho_{A}={e}^{-\mathcal{H}_{A}}$.
$|\alpha\rangle_{A, B}$ are the orthogonal Schmidt basis.
By using the variational method~\cite{RevModPhys.77.259, SCHOLLWOCK201196}, we can obtain a ground state of the corresponding Hamiltonian iteratively.
We have tested that both random initial MPS and the ``infinite'' method to initialize the system provide accurate ground states.
An generic and effective numerical criterion to estimate how accurately the wavefunction is approximated by an MPS with a fixed bond dimension is the so-called \emph{variance}~\cite{McCulloch_2007, PhysRevB.97.045125}
\begin{equation}
    v^{2}
    =\langle\psi|H^{2}-E_{0}^{2}|\psi\rangle,
    \label{eq:}
\end{equation}
which is easy to compute in the MPS-MPO framework~\cite{SCHOLLWOCK201196}.
It is known that $v$ is proportional to the truncation error in the density matrix renormalization group (DMRG) method~\cite{McCulloch_2007}. Moreover $v$ is essentially determined by the distribution of the entanglement spectrum~\cite{PhysRevA.78.032329}.
For a gapped state, it tends to be exponentially small given a sufficient but not too large $\chi$.
When the system is approaching criticality and becoming long-range entangled, $v$ would increase rapidly since the correlation length $\xi$ becomes as large as the system size and the entanglement spectrum is much more densely distributed, resulting in a much larger entanglement entropy (EE) $S_{A}=-\text{tr}\left(\rho_{A}\ln\rho_{A}\right)\propto\ln\xi$ in comparison to gapped ones~\cite{Calabrese_2004, PhysRevLett.102.255701}.
Variance can be used to distinguish different phases and identify the critical points between them.

Because of the translational symmetry breaking by the OBC on a finite lattice, the open ends can induce dimerization and hence oscillations in bond energy and EE in 1d quantum many-body systems~\cite{PhysRevLett.96.100603, Affleck_2009}.
Therefore, the bipartite EE for subsystem $A$ has the following form~\cite{PhysRevB.99.205153}
\begin{equation}
    S_{A}(l)
    =S_{A}^{u}(l)+(-)^{l}S_{A}^{o}(l)+S_{0},
    \label{eq:entanglement_entropy_parts}
\end{equation}
where $l=0, \dots, L-2$ denotes the links.
$S_{A}^{u}(l)$ is the uniform part and
$S_{A}^{o}(l)$ is the oscillation part of the EE.
$S_{A}^{o}(l)$ is phenomenologically proportional to the oscillatory part of the bond energy as $S_{A}^{o}(l)=\alpha E^{o}(l)$~\cite{PhysRevLett.96.100603}.
The bond energy behaves as $E_{b}(l)\equiv\langle{h_{l}}\rangle=E_{b}^{u}+(-)^{l}E_{b}^{o}(l)$ where $E_{b}^{u}$ is a constant, which can be extracted as $E_{b}^{u}=\frac{1}{2}\left[E_{b}(L/2)+E_{b}(L/2+1)\right]$.
Once $S_{A}^{u}(l)$ is extracted by fitting and finding the optimal $\alpha$,
on a finite lattice with the number of bonds $\mathcal{L}=L-1$, Cardy's formula under OBC reads~\cite{Calabrese_2004}
\begin{equation}
    S_{A}^{u}(l)
    =\frac{c}{6}\ln\left[\frac{2\mathcal{L}}{\pi}\sin\frac{\pi(l+1/2)}{\mathcal{L}}\right]+S_{0},
    \label{eq:cardy_formula}
\end{equation}
where $c$ is the central charge characterizing the corresponding conformal field theory (CFT), and $l=0, \dots, \mathcal{L}-1$.
However, in a 1d chain consisting of an odd number of sites, EE will develop plateaus stemming from the incommensurate oscillations~\cite{PhysRevB.87.094415} and exact-zero modes in our LSM system, which may cause an underestimated central charge.
We demonstrate this issue with a simpler example, the 1d XY model, in the Appendix~\ref{sec:appdx_A}.

\section{The phase diagram \label{sec:phase_diagram}}

Before discussing the phase diagram of the 1d system, we first restrict the phase space to be studied by symmetry analysis. In the fermion model Eq.~(\ref{eq:fermion_ham}) parametrized by $(t,\Delta,\Delta^\prime,V)$, it is straightforward to verify the following symmetries:
\bea\notag
&H^\ast(t,\Delta,\Delta^\prime,V)=H(t,\Delta,-\Delta^\prime,V),\\
\notag&T_xH(t,\Delta,\Delta^\prime,V)T_x^{-1}=H(-t,-\Delta,\Delta^\prime,V),\\
\notag&e^{\imth\frac\pi2\sum_r\hat n_r}H(t,\Delta,\Delta^\prime,V)e^{-\imth\frac\pi2\sum_r\hat n_r}=H(t,-\Delta,-\Delta^\prime,V).
\eea
In other words, changing the sign of hopping $t$, or real pairing $\Delta$, or imaginary pairing $\Delta^\prime$ does not affect the spectrum of the 1d chain, in the thermodynamic limit. Therefore, we set $t=1$ to be a positive constant, and restrict our numerical studies to the parameter regime $\Delta>0,~\Delta^\prime>0$. Below we present our results on the phase diagram of 1d model as Eq.~(\ref{eq:fermion_ham}) or (\ref{eq:spin_ham}).

\subsection{The non-interacting limit}

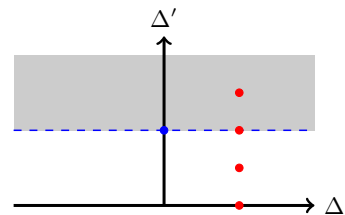
\begin{figure}[hbt!]
    \centering
    % \begin{tikzpicture}[x=1cm, y=1cm]
    \begin{tikzpicture}[]
        \pgfmathsetmacro\de{0.0};
        \pgfmathsetmacro\xshift{1.5*sqrt(3)};
        \begin{scope}
        \path[fill=gray!40] (-2.0, 1.0) rectangle (2.0, 2.0);
        % \path[fill=gray!40] (-2.0, -1.0) rectangle (2.0, -2.0);
        \draw[->, line width=0.4mm] (-2.0, 0) -- (2.0, 0) node[right] {$\Delta$};
        \draw[->, line width=0.4mm] (0.0, -0.0) -- (0, 2.25) node[above] {$\Delta^{\prime}$};
        % \draw[dashed, line width=0.4mm, color=blue] (0.0, -1.0) -- (0.0, 1.0);
        \draw[dashed, line width=0.2mm, color=blue] (-2.0, 1.0) -- (2.0, 1.0);
        % \draw[dashed, line width=0.2mm, color=blue] (-2.0, -1.0) -- (2.0, -1.0);
        \draw[fill, color=blue] (0.0, 1.0) circle (0.5mm);
        % \draw[fill, color=blue] (0.0, -1.0) circle (0.5mm);
        \draw[fill, color=red] (1.0, 0.0) circle (0.5mm);
        \draw[fill, color=red] (1.0, 0.5) circle (0.5mm);
        \draw[fill, color=red] (1.0, 1.0) circle (0.5mm);
        \draw[fill, color=red] (1.0, 1.5) circle (0.5mm);
        \end{scope}
    \end{tikzpicture}
    \caption{Phase diagram for Eq.~(\ref{eq:fermion_ham}) with $V=0$. Gray area denotes gapless phases whereas white areas denote the gapped Kitaev chain. The axis perpendicular to $(\Delta, \Delta^{\prime})$ is to add the non-vanishing interaction $V$.}
   \label{fig:noninterating_pd}
\end{figure}

If $V=0$, Eq.~(\ref{eq:fermion_ham}) is free-fermion model that can be solved exactly, and it provides a good starting point to understand the full phase diagram of the interacting model.
We consider a closed 1d chain of $\mathcal{N}=2N$ sites (or $N$ unit cells) Under periodic boundary condition (PBC), where we label the sites by $j\in\{0, \dots, \mathcal{N}-1\}$. Note that there are two sites $j=0, \mathcal{N}/2$ as inversion centers in this 1d chain. There are $N=\mathcal{N}/2$ unit cells, labeled as $l=0,\cdots,N-1$, with the site index $j=2l+\alpha$, $l=0, \dots, N-1$, $\alpha=0, 1$.
Therefore the free fermion Hamiltonian can be rewritten as
\begin{equation}
    \begin{aligned}
        H_{0}&
        =\sum_{\alpha, l}\left[(-)^{\alpha}t c_{2l+\alpha}^{\dagger}c_{2l+\alpha+1}+h.c.\right] \\
        &+\sum_{\alpha, l}\left(\left[(-)^{\alpha}\Delta+\text{i}\Delta^{\prime}\right]c_{2l+\alpha}^{\dagger}c_{2l+\alpha+1}^{\dagger}+h.c.\right).
    \end{aligned}
\end{equation}
Fourier transformations are defined as
\begin{equation}
    d_{k, \alpha}=\frac{1}{\sqrt{N}}\sum_{l=0}^{N-1}e^{\text{i}kl}c_{2l+\alpha}
\end{equation}
In the spinor basis of $\eta_{k}=(d_{k, 0}, d_{-k, 0}^{\dagger}, d_{k, 1}, d_{-k, 1}^{\dagger})^{T}$, $H_{0}$ has the following form in momentum space
\begin{equation}
H_{0}=\sum_{k\geqslant 0}\eta_{k}^{\dagger}\Gamma(k)\eta_{k},
\end{equation}
with
\begin{equation}
\Gamma(k)=
\begin{pmatrix}
 0 & 0 & t(k) & \Delta(k) \\
 0 & 0 & -\Delta^{*}(-k) & -t(k) \\
 t^{*}(k) & -\Delta(-k) & 0 & 0 \\
 \Delta^{*}(k) & -t^{*}(k) & 0 & 0
\end{pmatrix},
\end{equation}
where we define $t(k)\equiv t(1-e^{\text{i}k})=-2\text{i}t\sin\left(\frac{k}{2}\right)e^{\text{i}k/2}$ and $\Delta(k)\equiv\left[\Delta\left(1+e^{\text{i}k}\right)+\text{i}\Delta^{\prime}\left(1-e^{\text{i}k}\right)\right]=2\left[\Delta\cos\left(\frac{k}{2}\right)+\Delta^{\prime}\sin\left(\frac{k}{2}\right)\right]e^{\text{i}k/2}$.
Note that $t(-k)=t^{*}(k)$. We can obtain the four-band dispersion relations as $\pm\epsilon(\pm k)$, where
\begin{equation}
\epsilon(k)=2\left[\sqrt{t^{2}\sin^{2}\left(\frac{k}{2}\right)+\Delta^{2}\cos^{2}\left(\frac{k}{2}\right)}
+\Delta^{\prime}\sin\left(\frac{k}{2}\right)\right].
\end{equation}
We denote the two positive eigenvalues as $\epsilon_{0}(k)\geqslant\epsilon_{1}(k)\geqslant 0,~\forall{k}\geqslant 0$.
Therefore the Hamiltonian will be diagonalized to the form $H_{0}=\sum_{k\geqslant 0}\gamma_{k}^{\dagger}\Lambda(k)\gamma_{k}$, in which $\Lambda(k)=\text{diag}\left\{\epsilon_{0}(k), -\epsilon_{0}(k), \epsilon_{1}(k), -\epsilon_{1}(k)\right\}, \gamma_{k}=(f_{k, 0}, f_{-k, 0}^{\dagger}, f_{k, 1}, f_{-k, 1}^{\dagger})^{T}$.
\begin{comment}
In another explicit form,
\begin{equation}
    \begin{aligned}
        H_{0}&=\sum_{k\geqslant 0}\left[\epsilon_{0}(k)f_{k, 0}^{\dagger}f_{k, 0}+\epsilon_{0}(k)f_{-k, 0}^{\dagger}f_{-k, 0}-\epsilon_{0}(k)\right] \\
        &+\sum_{k\geqslant 0}\left[\epsilon_{1}(k)f_{k, 1}^{\dagger}f_{k, 1}+\epsilon_{1}(k)f_{-k, 1}^{\dagger}f_{-k, 1}-\epsilon_{1}(k)\right].
    \end{aligned}
\end{equation}
\end{comment}
\begin{comment}
If we extend the definition of $\epsilon_{0, 1}(k), k\geqslant 0$ to the original domain of definition $(-\pi, \pi]$ by $\epsilon_{0, 1}(-k)\equiv\epsilon_{0, 1}(k)$, then we have
\begin{equation}
H_{0}=\sum_{\sigma=0, 1}\sum_{k}\epsilon_{\sigma}(k)\left(f_{k, \sigma}^{\dagger}f_{k, \sigma}-\frac{1}{2}\right).
\end{equation}
$\sigma=0, 1$ denote the two branches of fermions.
$\epsilon_{\sigma}(k)\geqslant 0$ is the quasi-particle excitation spectrum.
The negative Dirac sea is filled.
The ground state energy density is given by
$\varepsilon_{0}=\frac{1}{2N}\sum_{\sigma, k}\left[-\frac{1}{2}\epsilon_{\sigma}(k)\right]
=-\frac{1}{8\pi}\sum_{\sigma, k}\left(\frac{2\pi}{N}\right)\epsilon_{\sigma}(k)\xrightarrow{\lim_{N\rightarrow\infty}}-\frac{1}{8\pi}\sum_{\sigma}\int_{-\pi}^{\pi}dk\epsilon_{\sigma}(k)$.
\end{comment}
If we set $t=1.0$ as the energy unit, the non-interacting phase diagram $(\Delta, \Delta^{\prime})$ for $V=0$ is illustrated in FIG.~\ref{fig:noninterating_pd}.
Some representative cases are discussed as follows:
\begin{itemize}
\item
$\Delta=0, \Delta^{\prime}=1.0$ is a special case where the lower band is flat lying exactly at zero energy. It means that gapless excitations appear for all $k$.
If $\Delta^{\prime}\neq{1.0}$, the system features two linearly dispersing Majorana modes at $k_{0}=0$: they have different velocities and are hence not conformally invariant.
\item
$\Delta\neq 0, \Delta^{\prime}<1.0$ always give us a gapped superconductor, i.e. a Majorana chain.
Particularly, $\Delta=1.0, \Delta^{\prime}=0.0$ features a flat band spectrum of Bogoliubov quasiparticles. 
\item
$\Delta\neq 0, \Delta^{\prime}=1.0$ is gapless at $k_{0}=\pi$.
The dispersion is expanded as $\pm\epsilon_{1}(k_{0}+\delta k)\approx\pm\frac{1}{4}\delta k^{2}+\mathcal{O}(\delta k^{2})$, which is quadratic near $k_0=\pi$.
\item
$\Delta\neq 0, \Delta^{\prime}>1.0$ gives rise to a gapless phase with a pair of linearly dispersing Majorana modes.
The gapless point is located at $k_{0}=2\arctan\frac{\Delta}{\sqrt{\Delta^{\prime 2}-t^{2}}}$.
Around this point, the dispersion relation reads $\pm\epsilon_{1}(k_{0}+\delta k)\approx\pm{v_{s}}(k_{0})\delta{k}+\mathcal{O}(\delta k^{2})$, which is linear.
The speed of the Majorana mode is
\begin{equation}
v_{s}(k_{0})=\frac{t^{2}-\Delta^{2}-\Delta^{\prime 2}}{\Delta^{\prime}}\cos\left(\frac{k_{0}}{2}\right).
\end{equation}
\end{itemize}

\subsection{A representative study for the interacting cases: $\Delta=1.0$}

By fixing $\Delta=1.0$ and selecting different $\Delta^{\prime}$, we can obtain a rough idea of the whole phase diagram for the interacting model Eq.~(\ref{eq:fermion_ham}).
First we choose a fixed system size $L=121$, which we find is large enough to accurately illustrate the phase diagram.
Note that, in the main text, the bond dimension of all MPSs is fixed at $\chi=64$, which we find is sufficient to obtain converged physical measurements.
A larger bond dimension is tested and verified in Appendix~\ref{sec:appdx_B}.
Furthermore, we select $\Delta^{\prime}=0.0, 0.5$, which belong to gapped superconductor phases at $V=0$; and $\Delta^{\prime}=1.0, 1.5$, which are gapless at $V=0$ as we discussed above in the non-interacting limit.
These four cases are illustrated by red dots in FIG.~\ref{fig:noninterating_pd}.
For $V\rightarrow\pm\infty$, obviously Eq.~(\ref{eq:spin_ham}) will lead to  antiferromagnetic$-z$ (ferromagnetic$-z$) states, respectively.
Therefore, we restrict ourselves to scan the parameter range ${-15.0}\leqslant{V}\leqslant{15.0}$, which turns out to be sufficient.

\begin{figure}[hbt!]
     \centering
     \includegraphics[width=0.5\textwidth]{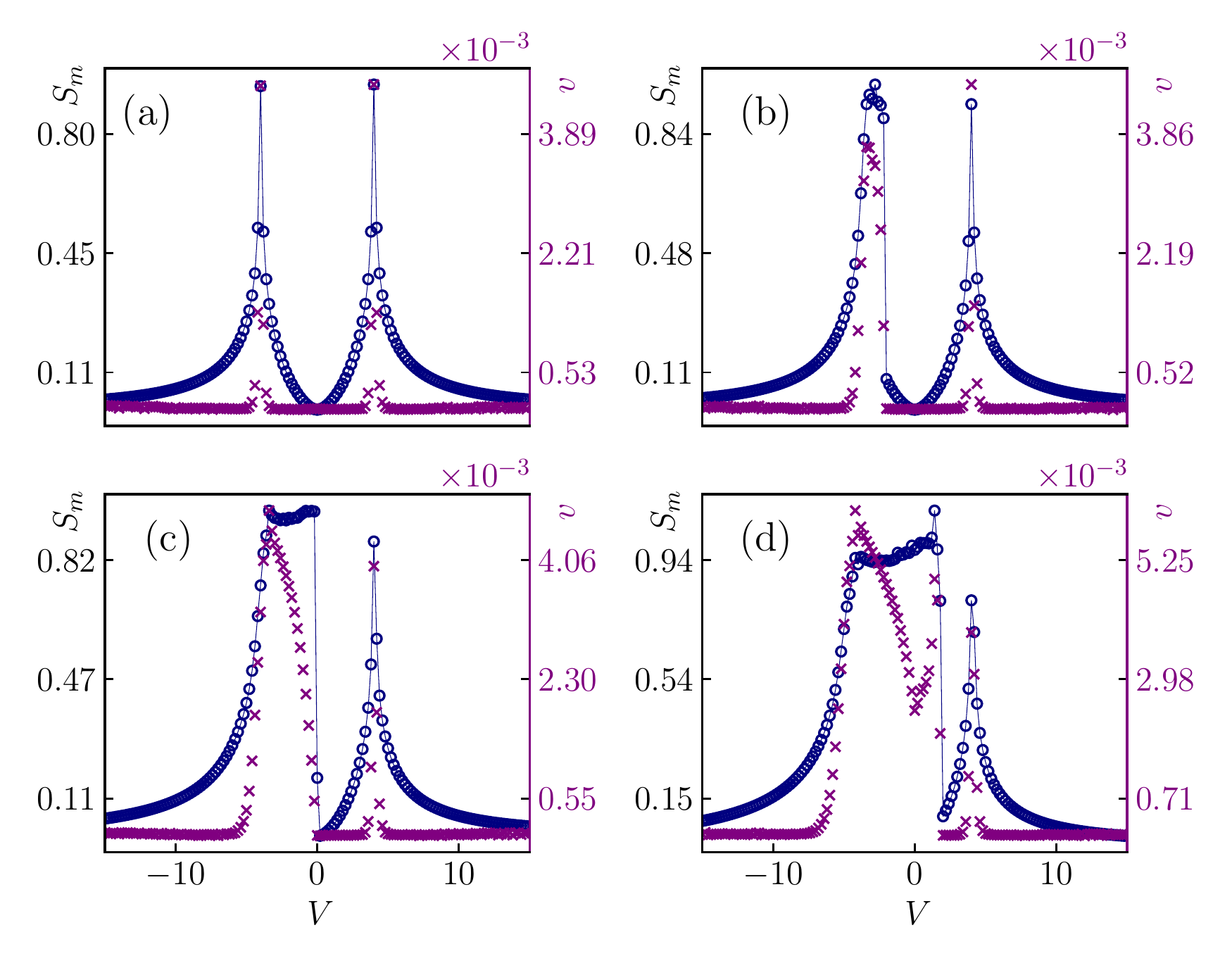}
     \caption{Middle-bond entanglement entropy $S_{m}$ (blue circle) and variance $v$ (purple cross) in the case of $\Delta=1.0$. $L=121$. (a, b, c, d) denote $\Delta^{\prime}=(0.0, 0.5, 1.0, 1.5)$, respectively.
     }
     \label{fig:flsm_maxee_variance_lattice121_chi64_d10}
\end{figure}

\begin{figure}[hbt!]
     \centering
     \includegraphics[width=0.49\textwidth]{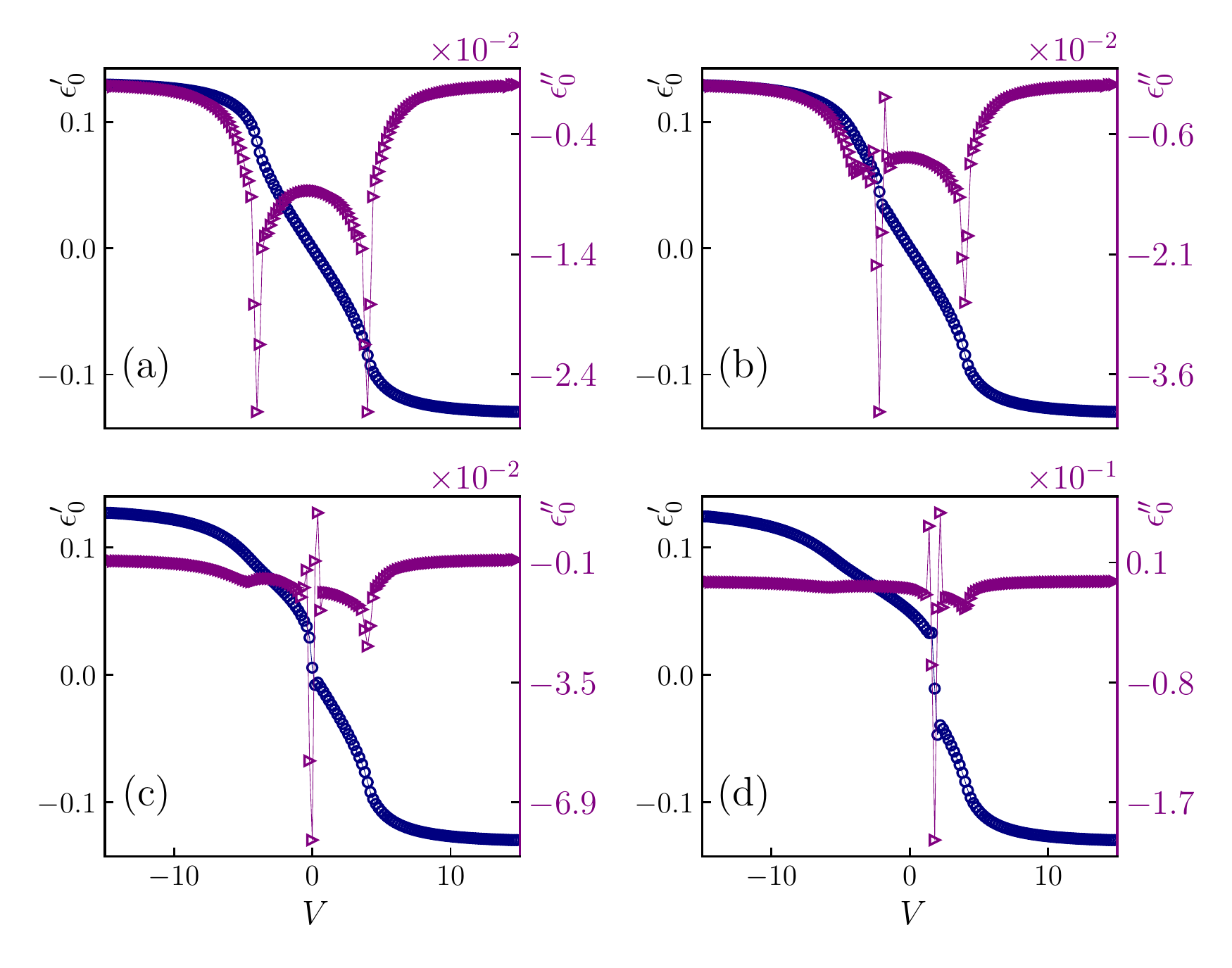}
     \caption{First (blue circle markers) and second (purple triangle markers) derivatives of the ground state energy density $\epsilon_{0}$. $\Delta=1.0$. $L=121$. (a, b, c, d) denote $\Delta^{\prime}=(0.0, 0.5, 1.0, 1.5)$, respectively.
     }
     \label{fig:flsm_ene_diff_lattice121_chi64_d10}
\end{figure}

\begin{figure}[hbt!]
     \centering
     \includegraphics[width=0.49\textwidth]{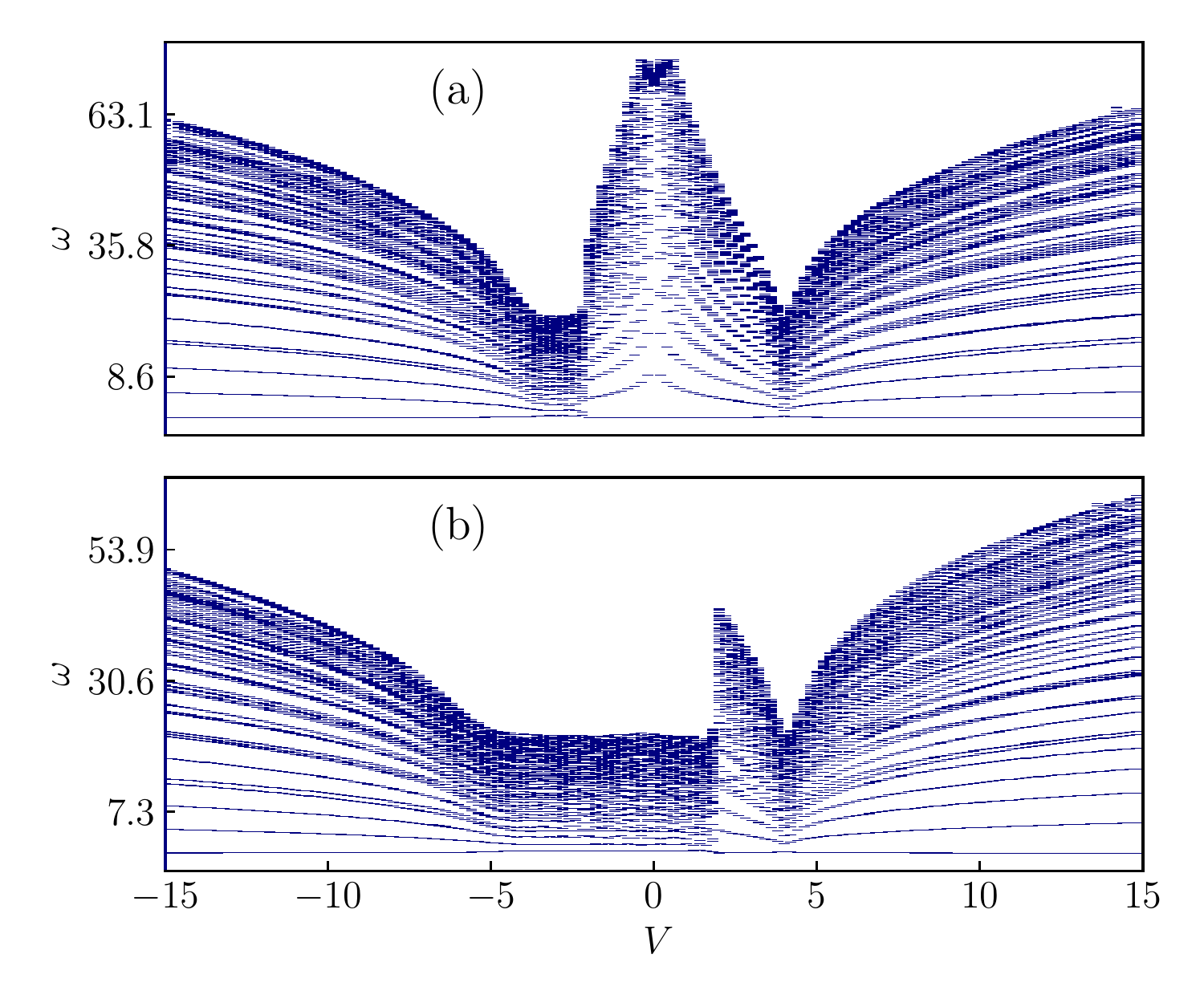}
     \caption{Middle-bond bipartite entanglement spectrum defined by Eq.~(\ref{eq:schmidt}). $\Delta=1.0$. $L=121$. (a) $\Delta^{\prime}=0.5$. (b) $\Delta^{\prime}=1.5$.
     }
     \label{fig:flsm_es_lattice121_d10}
\end{figure}

Once obtaining a converged ground state from a randomly initialized MPS, first we plot the middle-bond EE $S_{m}$ and variance $v$ in FIG.~\ref{fig:flsm_maxee_variance_lattice121_chi64_d10} as the functions of $V$, in which we can identify the gapped ground states with a vanishing variance $v$.
EE and variance maximize at the same time implying that the system is approaching a critical point.
For $V>0$, we notice that that there is a robust critical point at $V\approx{4.0}$, which seems independent of the choice of $\Delta^{\prime}$.
While for $V<0$, non-vanishing $\Delta^{\prime}$ can expand one critical point for $\Delta^{\prime}=0.0$ (see FIG.~\ref{fig:flsm_maxee_variance_lattice121_chi64_d10}(a)) to a finite gapless regime as shown in FIG.~\ref{fig:flsm_maxee_variance_lattice121_chi64_d10}(b, c, d).
Larger $\Delta^{\prime}$ induces a wider gapless phase.
For example, $\Delta^{\prime}=1.5$ can induce a critical phase for $-5.0\lessapprox{V}\lessapprox{2.0}$.
Within this gapless phase, the variance $v$ also fluctuates and shows distinctions between $V>0$ and $V<0$ regions, which implies that the gapless phases can be further distinguished.
By taking both FIG.~\ref{fig:flsm_ene_diff_lattice121_chi64_d10}(d) and FIG~\ref{fig:flsm_maxee_variance_lattice121_chi64_d10}(d) into consideration, we can conclude that the left phase boundary at $V\approx{-5.0}$ of this $\Delta^{\prime}$ induced gapless phase is of a quantum Berezinsky-Kosterlitz-Thouless (BKT)~\cite{Kosterlitz_1973} type, characterized by a central charge $c=1$.
The right phase boundary at $V\approx 2.0$ is of a first-order type.
Different types of quantum phases and their transitions can also be further illustrated by the distributions of bipartite entanglement spectrum in the ground state.
We use $\Delta^{\prime}=0.5, 1.5$ as two examples, which are shown in  FIG.~\ref{fig:flsm_es_lattice121_d10}.

\begin{figure}[hbt!]
     \centering
     \includegraphics[width=0.49\textwidth]{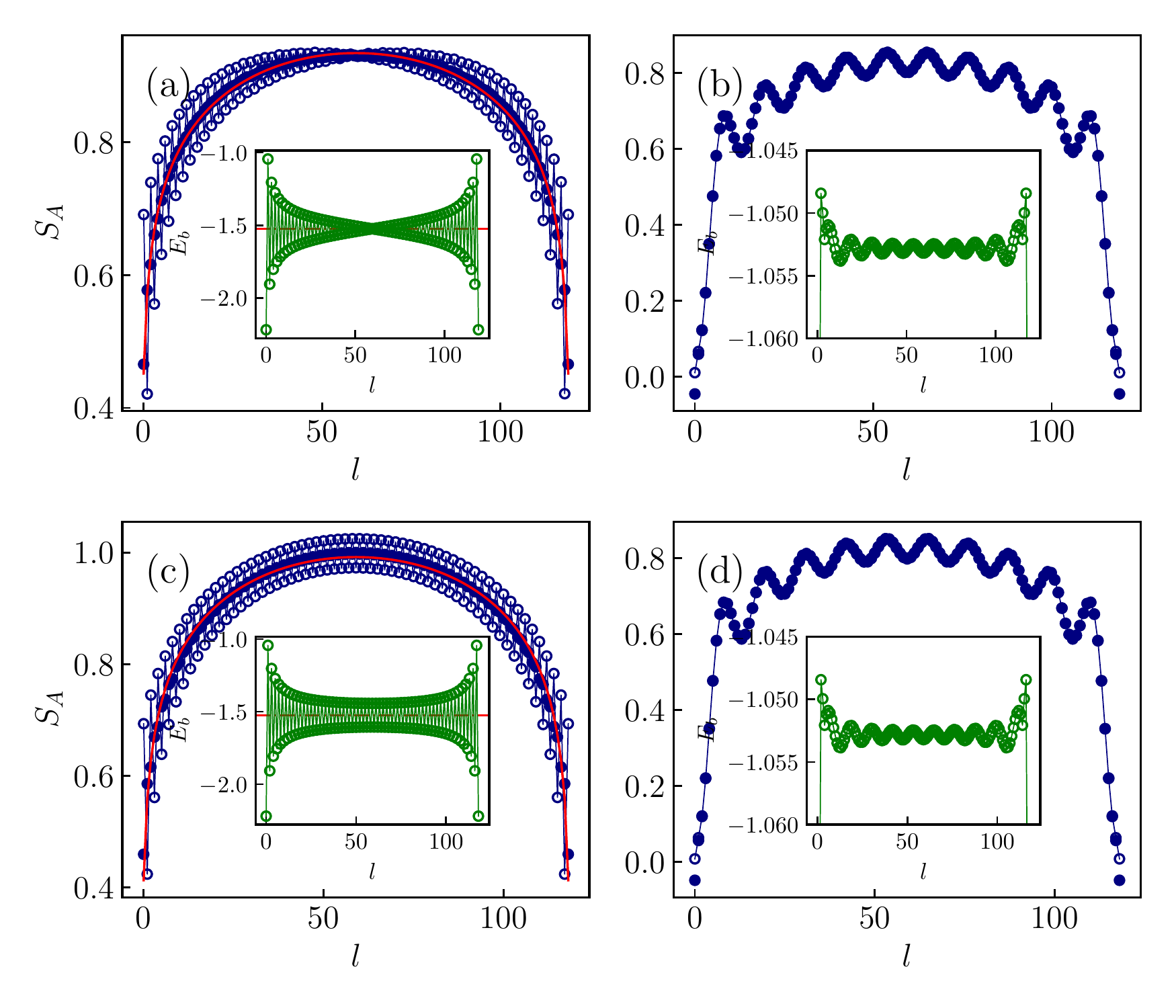}
     \caption{Bond energy $E_{b}$ (insets) and bipartite EE $S_{A}$. $\Delta=1.0, \Delta^{\prime}=1.5$. (a, b) $L=121$. (c, d) $L=120$. (a, c) $V=-3.0$. (b, d) $V=1.8$. Hollow circles represent the original data while filled circles represent the extracted uniform EE as defined in Eq.~(\ref{eq:entanglement_entropy_parts}).}
     \label{fig:flsm_ee_lattice121120_d10dp15}
\end{figure}

In FIG.~\ref{fig:flsm_ee_lattice121120_d10dp15}, we plot two examples of EE and bond energy for $\Delta=1.0, \Delta^{\prime}=1.5$.
FIG.~\ref{fig:flsm_ee_lattice121120_d10dp15}(a, c) denote $V=-3.0$, which is deep in the gapless phase.
By fitting from Cardy's formula Eq.~(\ref{eq:cardy_formula}),
the central charge reads $c\approx{0.574}$ for the odd lattice $L=121$, where we can see that the oscillations near the center of the 1d chain is incommensurate.
In comparison, in an even lattice of $L=120$, the central charge $c\approx{0.917}$ turns out to be quite different. We address this issue in Appendix \ref{sec:appdx_A} and attribute the underestimation of central charge on odd lattices to the exact ground state degeneracy, due to anti-commutation relation between inversion and parity symmetries.
Even lattices can be regarded as a perturbation which opens a small finite-size gap between the two degenerate ground states on an odd lattice, which leads to more accurate estimations of the central charge.
In FIG.~\ref{fig:flsm_ee_lattice121120_d10dp15}(b, d), $V=1.8$ lies in the gapless phase but in proximity to the first order transition into the Kitaev chain. Compared to $V=-3.0$ case in FIG.~\ref{fig:flsm_ee_lattice121120_d10dp15}(a, c),
here the oscillations in terms of both EE and bond energy are quite different, with a larger period persisting into the bulk. We found that fitting to Cardy's formula failed to produce a sensible central charge. 

\begin{figure}[hbt!]
     \centering
     \includegraphics[width=0.49\textwidth]{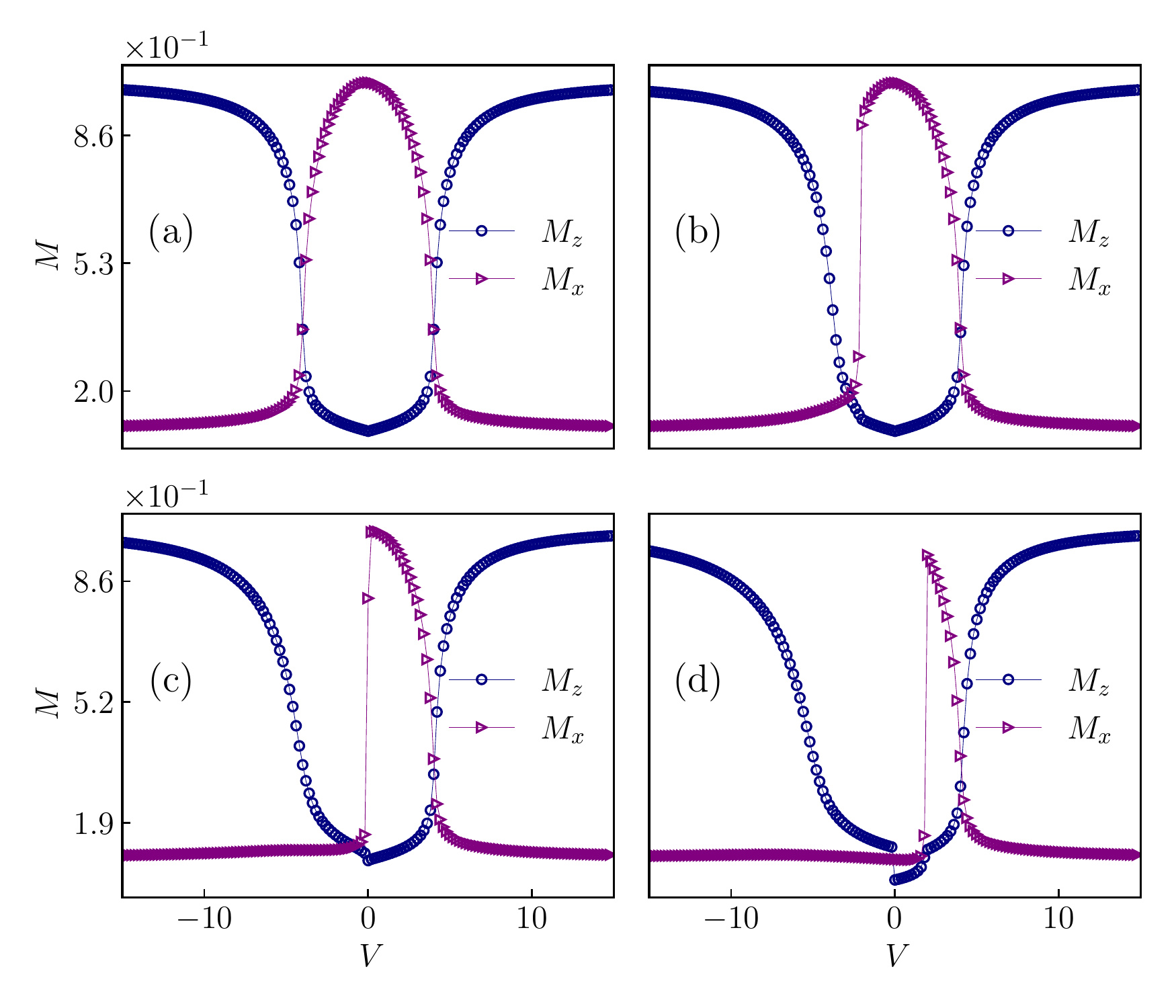}
     \caption{Magnetic order parameters $M_{z}$ and $M_{x}$ in the case of $\Delta=1.0$. $L=121$. For $V>0$, $M_{z}=M_{\text{AFM}-z}$. For $V<0$, $M_{z}=M_{\text{FM}-z}$. (a, b, c, d) denote $\Delta^{\prime}=(0.0, 0.5, 1.0, 1.5)$, respectively.
     }
     \label{fig:flsm_ops_lattice121_d10}
\end{figure}

In FIG.~\ref{fig:flsm_ops_lattice121_d10}, we plot the order parameters defined in Eq.~(\ref{eq:op_z}, \ref{eq:op_x}), from which we can see that, for $\Delta^{\prime}=1.5$, in the gapped phase $2.0\lessapprox{V}\lessapprox{4.0}$, the non-vanishing $\langle{M}_{x}\rangle$ implies the spontaneous breaking of parity symmetry $\mathcal{P}$, pointing to a Kitaev chain. Meanwhile for $V\rightarrow\pm\infty$, non-vanishing $\langle{M}_{z}\rangle$ implies breaking of inversion symmetry $\mathcal{I}$.
Because of their incompatible unbroken symmetries, the phase transition between these two symmetry breaking phases is beyond the Landau paradigm of spontaneous symmetry breaking. It requires a more detailed study, which we present in the next section.

\subsection{Overview of the whole phase diagram}

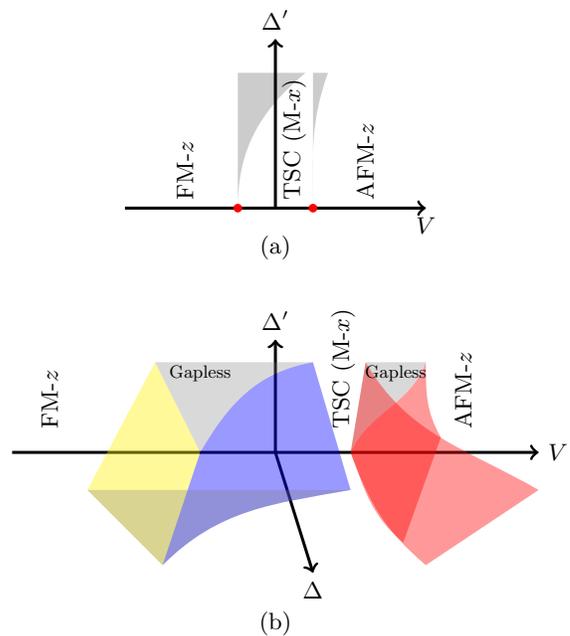
\begin{figure}[hbt!]
    \centering
    \begin{tikzpicture}[]
        \pgfmathsetmacro\de{0.0};
        \pgfmathsetmacro\xshift{1.5*sqrt(3)};
       \begin{scope}[shift={(0.0, 0.0)}]
            \path [fill=gray!40] (0.5, 0.0) to [out=90, in=250] (0.7, 1.8) -- (0.5, 1.8) -- (0.5, 0.0);
            \path [fill=gray!40] (-0.5, 0.0) to [out=90, in=220] (0.4, 1.8) -- (-0.5, 1.8) -- (-0.5, 0.0);
            \draw [->, line width=0.4mm] (-2.0, 0) -- (2.0, 0) node[below] {$V$};
            \draw [->, line width=0.4mm] (0, 0) -- (0, 2.25) node[above] {$\Delta^{\prime}$};
            \draw [fill, color=red] (-0.5, 0.0) circle (0.5mm);
            \draw [fill, color=red] (0.5, 0.0) circle (0.5mm);
            \node [anchor=west, rotate=90] at (0.25, 0.0) {TSC (M-$x$)};
            \node [anchor=west, rotate=90] at (1.2, 0.0) {AFM-$z$};
            \node [anchor=west, rotate=90] at (-1.2, 0.0) {FM-$z$};
            \node [anchor=north] at (0.0, -0.25) {(a)};
        \end{scope}

        \begin{scope}[shift={(0.0, -3.25)}]
        % \path[fill=gray!50] (-2.0, 1.0) rectangle (2.0, 2.0);
        % \path[fill=gray!50] (-2.0, -1.0) rectangle (2.0, -2.0);
        \draw[->, line width=0.4mm] (-3.5, 0) -- (3.5, 0) node[right] {$V$};
        \draw[->, line width=0.4mm] (0.0, 0.0) -- (0.0, 1.5) node[above] {$\Delta^{\prime}$};
        \draw[->, line width=0.4mm] (0.0, 0.0) -- (0.5, -1.6) node[below] {$\Delta$};
        % \path [fill=gray, opacity=0.3] (1.0, 0.0) to [out=290, in=140] (1.7, -1.2) to [out=130, in=260] (1.5, 0.0) -- (2.0, 1.2) -- (1.2, 1.2) -- (1.0, 0.0);
        \path [fill=gray, opacity=0.3] (1.2, 1.2) -- (2.0, 1.2) to [out=250, in=65] (1.0, 0.0);
        \path [fill=red, opacity=0.4] (1.0, 0.0) to [out=290, in=140] (2.0, -1.5) -- (3.5, -0.5) to [out=150, in=300] (1.2, 1.2) -- (1.0, 0.0);
        \path [fill=red, opacity=0.4] (1.0, 0.0) to [out=290, in=140] (1.7, -1.2) -- (2.2, 0.2) to [out=120, in=270] (2.0, 1.2) to [out=250, in=65] (1.0, 0.0);

        \path [fill=yellow, opacity=0.4] (-1.0, 0.0) -- (-1.5, -1.5) -- (-2.5, -0.5) -- (-1.6, 1.2) -- (-1.0, 0.0);
        \path [fill=blue, opacity=0.4] (-1.0, 0.0) -- (-1.5, -1.5) to [out=45, in=190] (1.0, -0.5) -- (0.5, 1.2) to [out=190, in=60] (-1.0, 0.0);
        \path [fill=gray, opacity=0.3] (-1.0, 0.0) to [out=60, in=190] (0.5, 1.2) -- (-1.6, 1.2);
        \path [fill=gray, opacity=0.3] (-1.5, -1.5) to [out=45, in=190] (1.0, -0.5) -- (-2.5, -0.5);

        \node [anchor=west, rotate=90] at (0.9, 0.2) {TSC (M-$x$)};
        \node [anchor=west, rotate=90] at (2.5, 0.2) {AFM-$z$};
        \node [anchor=west, rotate=90] at (-3.0, 0.2) {FM-$z$};
        \node [anchor=north, scale=0.75] at (1.6, 1.25) {Gapless};
        \node [anchor=north, scale=0.75] at (-1.0, 1.25) {Gapless};

        % \draw[dashed, line width=0.4mm, color=blue] (0.0, -1.0) -- (0.0, 1.0);
        % \draw[dashed, line width=0.2mm, color=blue] (-2.0, 1.0) -- (2.0, 1.0);
        % \draw[dashed, line width=0.2mm, color=blue] (-2.0, -1.0) -- (2.0, -1.0);
        % \draw[fill, color=blue] (0.0, 0.5) circle (0.5mm);
        \node [anchor=north] at (0.0, -2.0) {(b)};
        \end{scope}
    \end{tikzpicture}

    \caption{(a) Planar schematic phase diagram $(V, \Delta^{\prime})$ with $\Delta=1.0$. Gray areas denote gapless regimes. Red points denote DQCPs. (b) Three-dimensional schematic phase diagram $(V, \Delta, \Delta^{\prime})$ for the model. Red faces denote continuous second-order phase transitions. Blue face denotes discontinuous first-order phase transitions. Yellow face denotes BKT-type phase transitions. Closed faces form gapless regimes.}

   \label{fig:interacting_pd}
\end{figure}

We also obtained data for $\Delta=0.5, 1.5, 2.0$ in a range of different $\Delta^{\prime}$. The general structure of the phase diagrams as a function of $V$ are similar to $\Delta=1.0$ case. In particular, a larger $\Delta$ will drive the critical point $V_c$ between the Kitaev chain ($\langle M_x\rangle\neq0$) and inversion-breaking superconductor ($\langle M_z\rangle\neq0$) to a larger value. based on these numerical results, we can qualitatively draw the schematic phase diagrams as shown in FIG.~\ref{fig:interacting_pd}: (a) for a fixed $\Delta=1.0$ and (b) a three-dimensional phase diagram as a function of $(V,\Delta,\Delta^\prime)$. 

\section{Characterizing the critical behaviors \label{sec:critical_behaivor}}

\subsection{Finite-size analysis of the critical points}

After having a basic understanding of the phase diagram, we take a closer look at the critical point separating the parity-breaking Kitaev chain and the inversion-breaking superconductor. As we have mentioned, since the two gapped phases are not related to each other by spontaneous symmetry breaking, this unconventional critical point is beyond the Ginzburg-Landau-Wilson paradigm. 

In this subsection, we focus on the case with $\Delta=1.0$. In the first place, we compute the Binder cumulant~\cite{Binder1981, PhysRevLett.47.693}
\begin{equation}
    U_{L}
    =\frac{\langle{M}^{4}\rangle}{\langle{M}^{2}\rangle^{2}}
    \label{eq:}
\end{equation}
around the unconventional critical point $V\approx{4.0}$.
It follows the finite-size scaling ansatz $U_{L}=g_{U}\left(|\delta|L^{1/\nu}\right)$, in which $g_{U}$ does not scale with $L$.
$\nu$ is the critical exponent for the correlation length $\xi=|\delta|^{-\nu}$, in which $\delta$ is the reduced interacting parameter defined as $\delta\equiv{V-V_{c}}$.
Since the function $g_U$ is independent of the finite lattice sizes at the critical point,  numerical data for $U_{L}$ given by different lattice sizes will intersect at the same point.

\begin{figure}[hbt]
     \centering
     \includegraphics[width=0.49\textwidth]{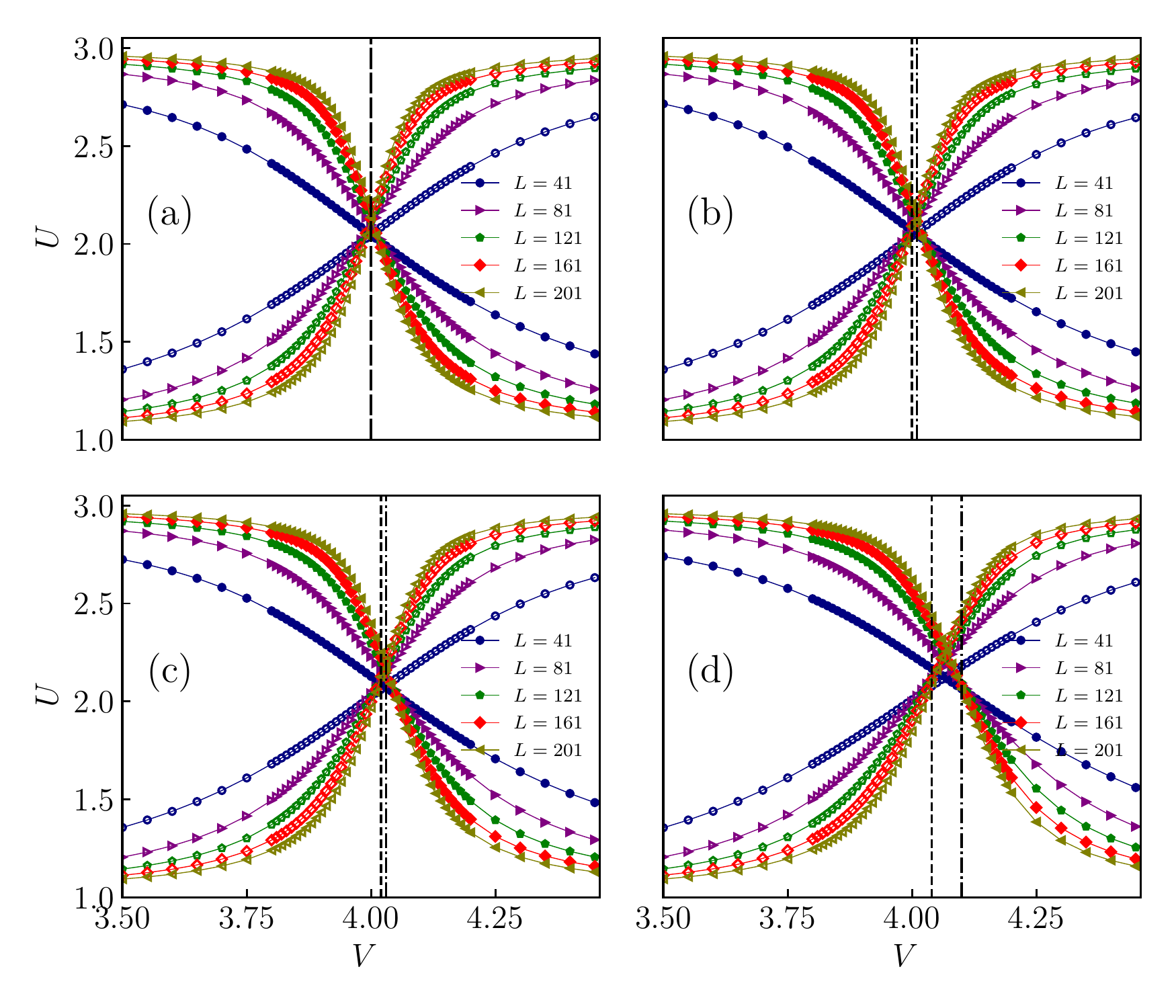}
     \caption{Binder cumulants $U_{z}$ (monotonically increasing solid markers) and $U_{x}$ (monotonically decreasing hollow markers) around the critical point in the case of $\Delta=1.0$. (a, b, c, d) denote $\Delta^{\prime}=(0.0, 0.5, 1.0, 1.5)$, respectively. Dashed lines mark the intersection point(s) for $U_{z}$ and $U_{x}$.
     }
     \label{fig:flsm_binder_odd_d10}
\end{figure}

In FIG.~\ref{fig:flsm_binder_odd_d10} we plot two Binder cumulants $U_{z}$ and $U_{x}$ corresponding to $M_{\text{AFM}-z}$ and $M_{x}$, respectively.
Furthermore, if we compute the derivative of the Binder cumulant, we can extract the correlation-length critical exponent $\nu$ since
\begin{equation}
\frac{dU}{dV}\propto{L}^{1/\nu}
\end{equation}
and it also reaches its maximum at the critical point~\cite{PhysRevX.5.041048}.

\begin{figure}[hbt!]
     \centering
     \includegraphics[width=0.49\textwidth]{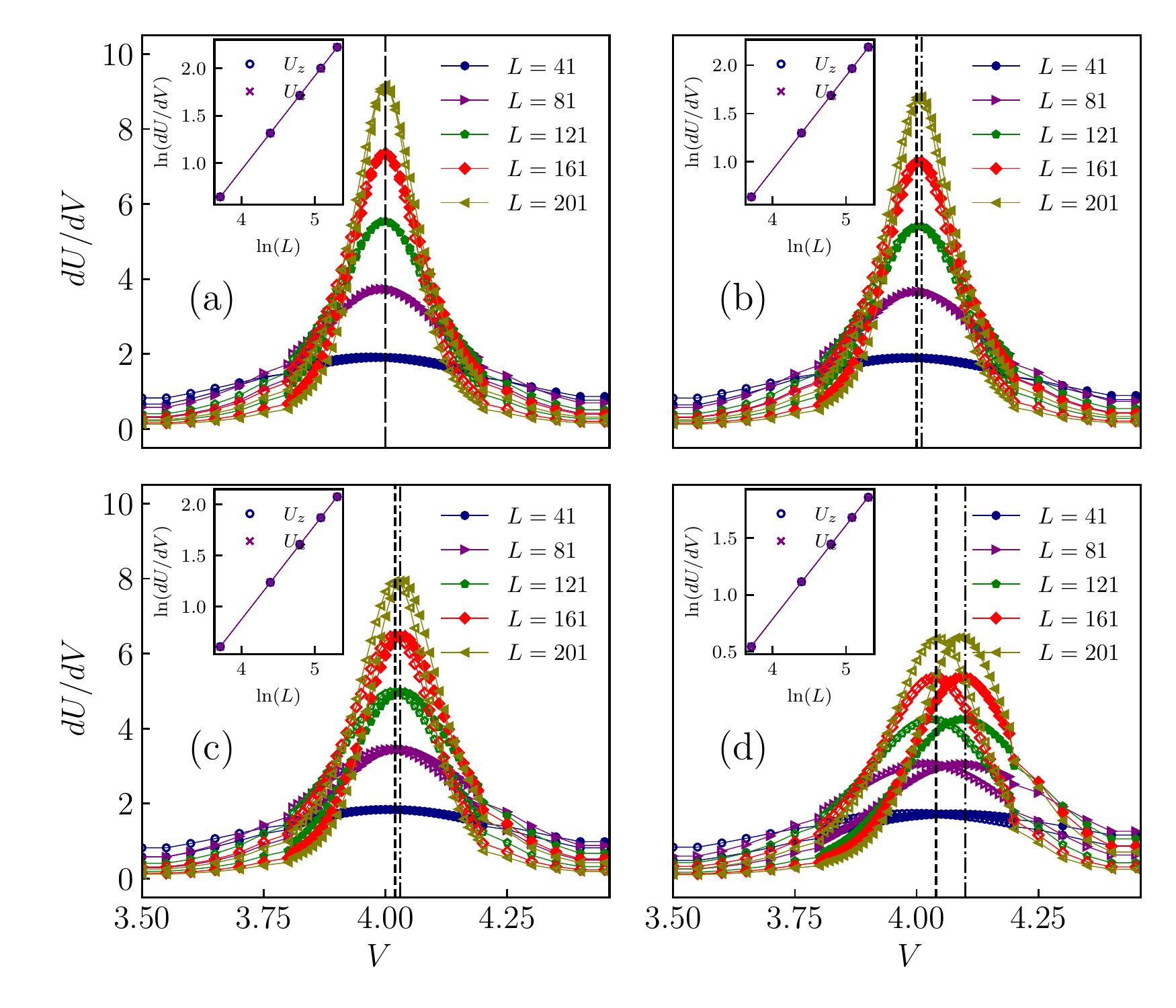}
     \caption{Derivatives of the Binder cumulants $dU_{z}/dV$ (solid markers) and $dU_{x}/dV$ (hollow markers) around the critical point. $\Delta=1.0$. (a, b, c, d) denote $\Delta^{\prime}=(0.0, 0.5, 1.0, 1.5)$, respectively. Dashed lines mark the peaks for $dU_{z}/dV$ and $dU_{x}/dV$. Insets show the finite-size logarithm fittings at the critical point(s) marked by the dashed lines.
     }
     \label{fig:flsm_binder_diff_odd_d10}
\end{figure}

\begin{figure}[hbt!]
    \centering
    \includegraphics[width=0.49\textwidth]{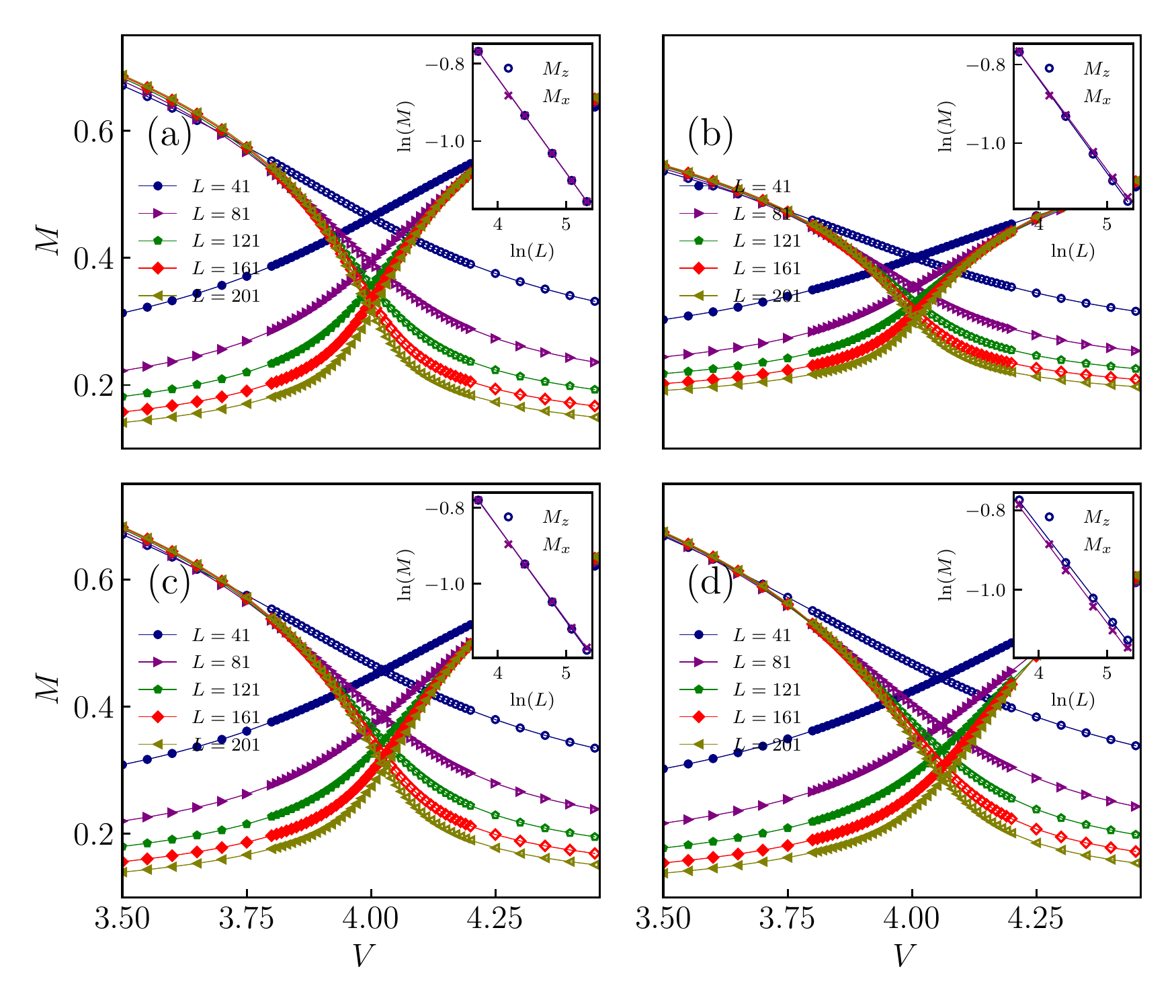}
    \caption{Magnetic order parameters $M_{\text{AFM}-z}$ (monotonically increasing solid data markers) and $M_{x}$ (monotonically decreasing hollow data markers) around the critical point. $\Delta=1.0$. (a, b, c, d) denote $\Delta^{\prime}=(0.0, 0.5, 1.0, 1.5)$, respectively. Insets show the logarithm finite-size fittings at the critical point(s).}
    \label{fig:flsm_mag_odd_d10}
\end{figure}

For $\Delta^{\prime}=0.0$ as shown in FIG.~\ref{fig:flsm_binder_odd_d10}(a), \ref{fig:flsm_binder_diff_odd_d10}(a), the two critical points determined by $U_{z}$ and $U_{x}$ coincide with each other, leading to a single critical point between the two gapped symmetry-breaking phases. This unique critical point is an analog of the deconfined quantum critical point (DQCP) in 2D~\cite{Senthil1490,PhysRevB.70.144407}, as we will discuss in more detail in the next subsection. As we gradually increase $\Delta^{\prime}>0$, we find that this single critical point starts to split into two, hosting a stable gapless phase in between. The two phase boundaries adjacent to it are determined by the scaling of two order parameters $M_x$ and $M_z$.
Furthermore, from the finite-size analysis in FIG.~\ref{fig:flsm_mag_odd_d10}, we can infer that both order parameters will vanish within this narrow gapless phase in the thermodynamic limit, suggesting that it preserves both inversion and parity symmetries.
This gapless nature of this phase is also inferred by a large central charge within it, which will be discussed in details later.

After nailing down the critical points, we turn to the finite-size scaling study for the corresponding magnetizations as
\begin{equation}
    \langle{M}\rangle=L^{-\beta/\nu}g_{M}\left(|\delta|L^{1/\nu}\right),
    \label{eq:}
\end{equation}
where $\beta$ is the critical exponent for the magnetic order parameter.
Therefore, exactly at the critical point we have $\langle{M}\rangle\propto{L^{-\beta/\nu}}$, which can be used to extract the related critical exponents.
This procedure is shown in FIG.~\ref{fig:flsm_mag_odd_d10} and insets there.
All these results are summarized in TABLE~\ref{tab:critical_exponents_d10}, where we can see that $M_{z, x}$ share the same critical exponents up to numerical errors even when the single critical point splits into two phase boundaries. This suggests an emergent symmetry relating the two order parameters. When the two order parameters become critical at the same point, this single critical point is a 1d DQCP as we will discuss soon. 

For a larger $\Delta$, we find that the single DQCP seems to persist for the full range of $\Delta^\prime$, instead of splitting into two phase boundaries with a gapless phase in between. We list the critical points and exponents for $\Delta=1.5$ in TABLE~\ref{tab:critical_exponents_d15}. Finally, we want to mention that in addition to the numerical error summarized in the Table, the finite step size $\delta V=0.02$ of data points can also lead to errors of the critical exponents, which is hard to evaluate. 
\begin{comment}
Note that, for $\Delta=1.5, \Delta^{\prime}=0.5$, the critical exponents for $M_{z, x}$ do not look like identical exactly.
We believe this very error comes from the numerical step-size issue since practically we find that it is extremely sensitive around the critical points when we move just one step even we choose a very small $\delta{V}=0.02$.
\end{comment}
% chi=64
\begin{table}[hbt!]
    \caption{\label{tab:critical_exponents_d10}Critical point(s) and critical exponents for $\Delta=1.0$.}
\begin{ruledtabular}
\begin{tabular}{cccccc}
    $\Delta^{\prime}$ & $\left[V_{c}^{z}, V_{c}^{x}\right]$ & $\nu_{z}$ & $\beta_{z}/\nu_{z}$ & $\nu_{x}$ & $\beta_{x}/\nu_{x}$ \\
    \hline
    0.0 & [4.00, 4.00] & 1.01(0) & 0.24(4) & 1.01(1) & 0.24(4) \\
    0.5 & [4.01, 4.00] & 1.02(7) & 0.23(9) & 1.02(9) & 0.23(4) \\
    1.0 & [4.03, 4.02] & 1.08(5) & 0.24(9) & 1.08(5) & 0.24(5) \\
    1.5 & [4.10, 4.03] & 1.20(1) & 0.22(3) & 1.21(1) & 0.22(8) \\
\end{tabular}
\end{ruledtabular}
\end{table}

% chi=64
\begin{table}[hbt!]
    \caption{\label{tab:critical_exponents_d15}Critical point(s) and critical exponents for $\Delta=1.5$.}
\begin{ruledtabular}
\begin{tabular}{cccccc}
    $\Delta^{\prime}$ & $\left[V_{c}^{z}, V_{c}^{x}\right]$ & $\nu_{z}$ & $\beta_{z}/\nu_{z}$ & $\nu_{x}$ & $\beta_{x}/\nu_{x}$ \\
    \hline
    0.0 & [5.00, 5.00] & 0.89(1) & 0.21(5) & 0.89(3) & 0.21(5) \\
    0.5 & [5.00, 5.00] & 0.89(7) & 0.23(2) & 0.91(2) & 0.20(3) \\
    1.0 & [5.02, 5.02] & 0.93(4) & 0.21(2) & 0.93(3) & 0.23(7) \\
    1.5 & [5.04, 5.04] & 0.99(7) & 0.23(5) & 1.00(3) & 0.24(2) \\
\end{tabular}
\end{ruledtabular}
\end{table}

\subsection{Realizing DQCPs by fusing two dualed order parameters and the possible breakdown of the Tomonaga-Luttinger liquid\label{sec:realize_DQCP}}

\begin{figure}[hbt!]
    \centering
    \includegraphics[width=0.49\textwidth]{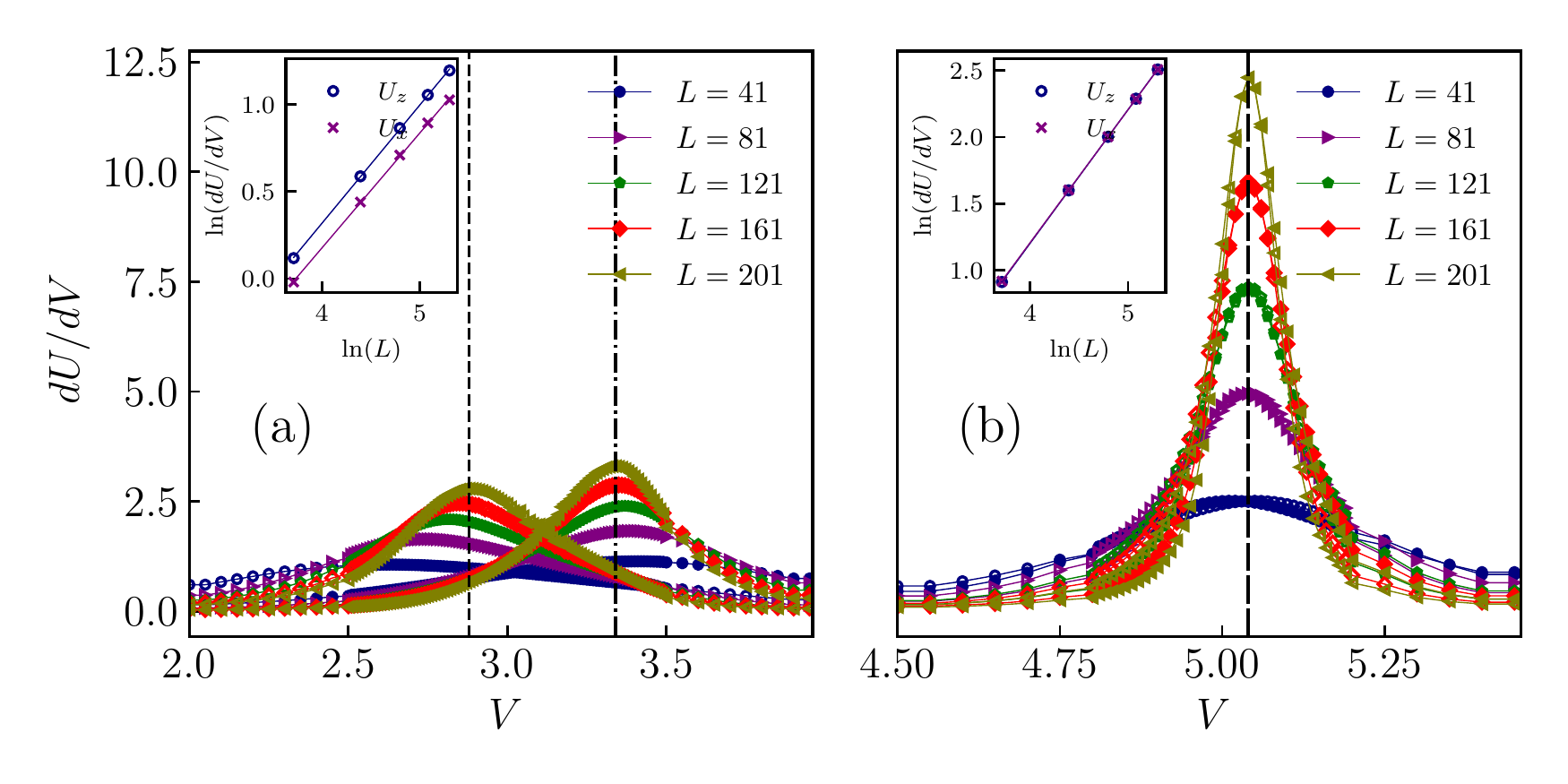}
     \caption{Derivatives of the Binder cumulants $dU_{z}/dV$ (solid markers) and $dU_{x}/dV$ (hollow markers) around the critical point(s). Insets show the logarithm finite-size fittings at the critical point(s). $\Delta^{\prime}=1.5$. (a) $\Delta=0.5$. (b) $\Delta=1.5$.
     }
     \label{fig:flsm_binder_diff_odd_d05d15}
\end{figure}

\begin{figure}[hbt!]
    \centering
    \includegraphics[width=0.49\textwidth]{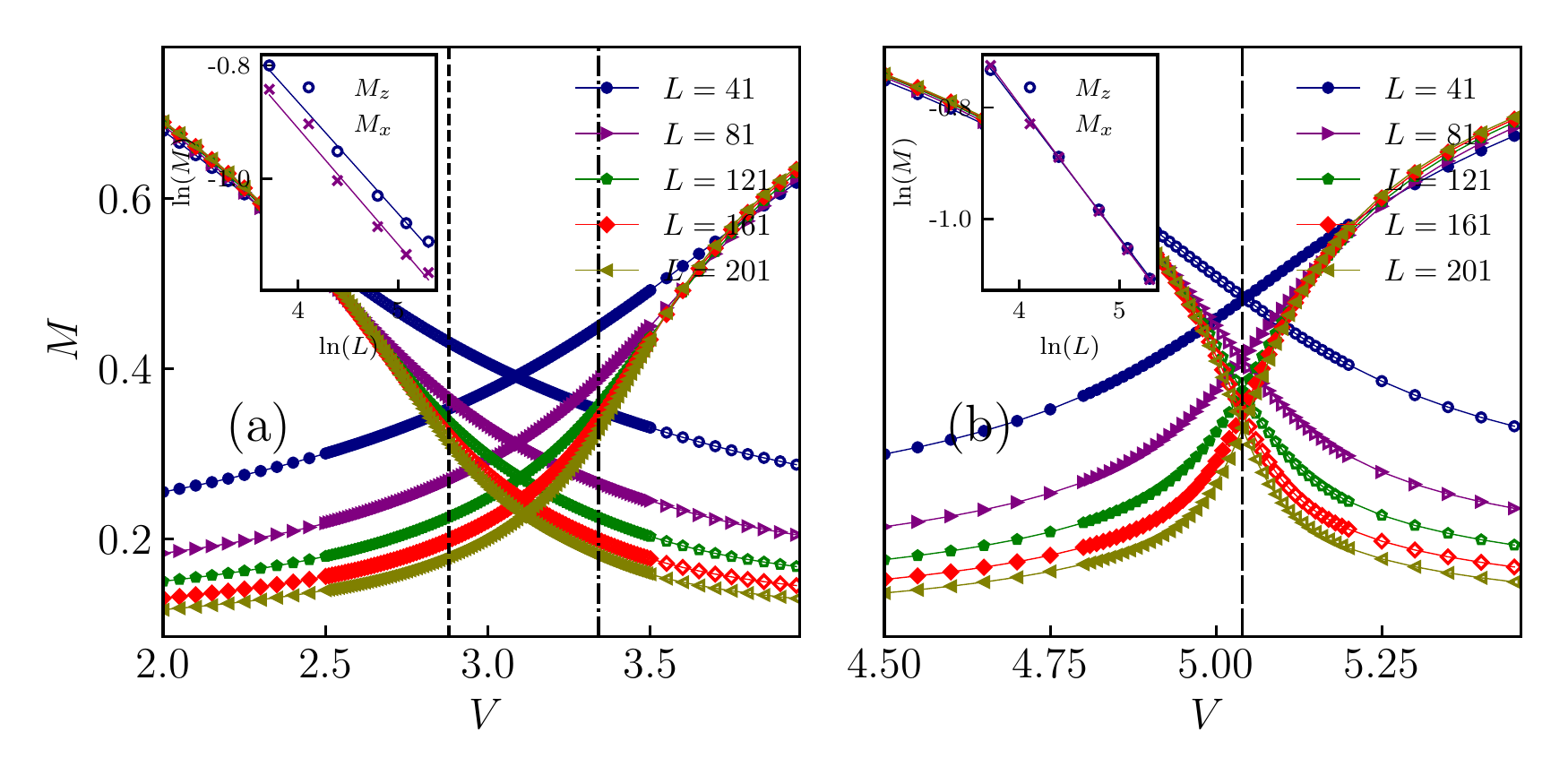}
     \caption{Magnetic order parameters $M_{\text{AFM}-z}$ (monotonically increasing solid markers) and $M_{x}$ (monotonically decreasing hollow markers) around the critical point. $\Delta^{\prime}=1.5$. (a) $\Delta=0.5$. (b) $\Delta=1.5$.
     }
     \label{fig:flsm_mag_odd_d05d15}
\end{figure}

We find that the parameter $\Delta$ determines the properties of the critical point(s) for $V>0$.
By fixing $\Delta^{\prime}=0.5$, we focus on two representative cases of $\Delta=0.5, 1.5$.
The finite-size scaling analysis similar to the previous section is shown in FIG.~\ref{fig:flsm_binder_diff_odd_d05d15} and FIG.~\ref{fig:flsm_mag_odd_d05d15}.
The critical exponents obtained from them are summarized in TABLE~\ref{tab:critical_exponents_d05d15}.

For $\Delta<1.0$, the two order parameters $M_{z, x}$ approach criticality at different $V$, leaving a narrow gapless phase between them where both order parameters $M_{z, x}$ vanish. Such a gapless phase is allowed by the LSM theorem. The peaks in FIG.~\ref{fig:flsm_binder_diff_odd_d05d15}(a) are also less sharp, which means the the corresponding critical exponent $\nu$ is larger.
However, for $\Delta>1.0$, $M_{z, x}$ approach criticality at one single critical point $V_c$.
The narrow gapless phase shrinks into a single critical point. The critical peaks in FIG.~\ref{fig:flsm_binder_diff_odd_d05d15}(b) are also sharper and lead to a smaller $\nu$.
In this sense, we say the two phase boundaries of vanishing $M_{z,x}$ fuse together and realize a DQCP.

To get more understanding for this process, we take a look at the effective field action of a U(1) Tomonaga-Luttinger liquid (TLL) theory expressed in terms of two variables $\phi$ and $\theta$:
\begin{equation}\label{eq:luttinger liquid}
    H_{LL}[\theta, \phi]=
    \frac{u}{2\pi}\int{dx}\left[\frac{1}{g}\left(\partial_{x}\theta\right)^{2}+g\left(\partial_{x}\phi\right)^{2}\right],
\end{equation}
where $u$ is the renormalized velocity and $g$ is the Luttinger parameter. Here $\Pi=\partial_x\theta/\pi$ is the canonical conjugate of phase variable $\phi$.

%In the dictionary of Abelian bosonization, the spin operators in a spin-$\frac12$ chain are expressed as
%\bea
%S^\pm(x)\sim e^{\mp\imth\theta}\big[(-1)^x\pm\cos2\phi\big],\\
%S^z(x)\sim-\frac{\nabla\phi}{\pi}+(-1)^x\cos2\phi.
%\eea
%Under the symmetry operations, the boson fields transform as
%\bea
%\mathcal{I}:~~&\theta(x)\rightarrow-\theta(-x),~\phi(x)\rightarrow\frac\pi2-\phi(-x),\\
%\mathcal{P}_f:~~&\theta(x)\rightarrow\theta(x)+\pi,~~\phi(x)\rightarrow\phi(x)\\
%\phs:~~&\theta(x)\rightarrow\theta(x),~\phi(x)\rightarrow\frac\pi2-\phi(x),\\
%\tilde{T}_x:~~&\theta(x)\rightarrow\pi-\theta(x+1),~\phi(x)\rightarrow\frac\pi2+\phi(x+1).~
%\eea
%where we have set the lattice constant as unity. Note that $\phs$ and $\tilde{T}_x$ are anti-unitary symmetries while inversion $\mathcal{I}$ and Ising $\mathcal{P}_f$ symmetries are unitary.

%The $\Delta=\Delta^\prime=0$ limit of our model (\ref{eq:spin_ham}) is nothing but the XXZ spin-$\frac12$ chain, which corresponds to[giamarchi]
%\bea
%&J_z/J_{x,y}=V/2t=-\cos^2(\pi g/2),\\
%&u(2-g)=\sqrt{(2t)^2-V^2}
%\eea
%for $|V|<2t$ in effective action (\ref{eq:luttinger liquid}). In particular, the Heisenberg limit $V=2t$ with $g=2$ and $u=\pi t$, characterizes the single critical point separating the Luttinger liquid phase at $|V|<2t$ and the Ising antiferromagnetic phase at $V>2t$.

In the dictionary of Abelian bosonization, the spin operators in a spin-$\frac12$ chain can be expressed as
\bea
S^z(x)\pm\imth S^x(x)\sim e^{\mp\imth\theta}\big[(-1)^x\pm\cos2\phi\big],\\
S^y(x)\sim-\frac{\nabla\phi}{\pi}+(-1)^x\cos2\phi.
\eea
Under the symmetry operations, the boson fields transform as
\bea
\mathcal{I}:~~&\theta(x)\rightarrow\pi-\theta(-x),~\phi(x)\rightarrow\frac\pi2-\phi(-x),\\
\mathcal{P}_f:~~&\theta(x)\rightarrow-\theta(x),~~\phi(x)\rightarrow\frac\pi2-\phi(x)\\
\phs:~~&\theta(x)\rightarrow\pi-\theta(x),~\phi(x)\rightarrow\phi(x),\\
\tilde{T}_x:~~&\theta(x)\rightarrow\theta(x+1),~\phi(x)\rightarrow\frac\pi2-\phi(x+1).~
\eea
where we have set the lattice constant as unity. Note that $\phs$ and $\tilde{T}_x$ are anti-unitary symmetries while inversion $\mathcal{I}$ and Ising $\mathcal{P}_f$ symmetries are unitary.

The $J_x=J_z$ (or equivalently $2(t+\Delta)=V$) limit of our model (\ref{eq:spin_ham}) corresponds to the XXZ spin-$\frac12$ chain, characterized by~\cite{giamarchi2004quantum}
\bea
&J_y/J_{x,z}=2(t-\Delta)/V=-\cos(\pi g/2),\\
&u(2-g)=\sqrt{V^2-4(t-\Delta)^2}=4\sqrt{t\Delta}
\eea
for $V\geq2|t-\Delta|$ in effective action (\ref{eq:luttinger liquid}). In particular, the Heisenberg limit $V=2t,~\Delta=0$ with $g=2$ and $u=\pi V$, characterizes the single critical point separating the Luttinger liquid phase at $V>2(t-\Delta)$ and the Ising antiferromagnetic phase at $V<2(t-\Delta)$.

The leading-order back-scattering terms introduced by $|J_x-J_z|$, $J_y$ and $\Gamma$ couplings in (\ref{eq:spin_ham}) are
\bea
\mathcal{H}_{b.s.}=g_\theta\cos(2\theta)+g_y\cos(4\phi)+g_\Gamma\nabla\phi\nabla\theta\cos\theta+\cdots
\eea
In particular the $g_\Gamma$ term from the $\Gamma$ (or $\Delta^\prime$) coupling in the lattice model has a scaling dimension of $\text{dim}(g_\Gamma)=2+\frac{g}4 $ and is hence irrelevant, suggesting the stability of the single critical point for a small $\Delta^\prime$ in the $\Delta=0$ limit. While the dimension of $g_y$ term is $\text{dim}(\cos4\phi)=4/g>2$, the $g_\theta$ term has a scaling dimension of $\text{dim}(\cos2\theta)=g<2$. Since the only relevant term is $\cos(2\theta)$, when $g_\theta$ changes sign, the ground state goes through a transition from the $AFM_z$ phase to the $AFM_x$ phase.

In this Luttinger liquid phase of model (\ref{eq:spin_ham}), the antiferromagnetic order parameters for the $AFM_{z,x}$ phases in the bosonized language read
\bea
M_z=\sum_j(-1)^jS^z_j\sim\cos\theta,\\
M_x=\sum_j(-1)^jS^x_j\sim\sin\theta,
\eea
and they share the same scaling dimension
\bea
\text{dim}[M_x]=\text{dim}[M_z]=\frac g{4}.
\label{eq:llt_sd}
\eea
We notice that a similar unification of two order parameters ($FM_z$ and VBS) at a DQCP in 1D is recently discussed by Ref.~\cite{PhysRevB.99.075103}. While in the context of Ref.~\cite{PhysRevB.99.075103} the relation between the two order parameters is only clear in the bosonized dual picture, here in our example of 1D DQCP a standard bosonization treatment already reveals the emergent symmetry between the two distinct order parameters $M_x$ and $M_z$, one ($M_z$) breaking inversion symmetry $\mathcal{I}$ while the other ($M_x$) breaks fermion parity $\mathcal{P}_f$.

The nature of this DQCP can be revealed by looking into the dual domain wall variables of e.g. the $AFM_z$ phase.
Here we follow the strategy of Ref.~\cite{PhysRevB.99.075103} to identify the projective symmetry action on the domain wall variables $\{\vec\mu_{j+1/2}|j\in\mathbb{Z}\}$ on the spin chain (\ref{eq:spin_ham}):
\bea
\notag&\mu_{j+1/2}^x=\sigma_j^z\sigma^z_{j+1},\\
\notag&\mu^z_{j-1/2}\rho_j^z\mu^z_{j+1/2}=\sigma_j^x,\\
&\rho_j^x=\sigma_j^z\label{eq:dual domain wall}
\eea
which are constraint by the Gauss' law
\bea
\mu^x_{j+1/2}=\rho^x_j\rho^x_{j+1}
\eea
Here $\rho_j^z$ is the link variable for the $Z_2$ gauge field, while $\mu^z_{j+1/2}$ creates a $Z_2$ gauge charge (i.e. domain wall of $M_z$ order parameter) on the dual site $j+\frac12$. In terms of the dual variables, the symmetry operations can be written as
\bea\label{sym:spin:ising:dual}
&\mathcal{P}_f=\prod_j\rho_j^x=\prod_{j=\text{odd}}\mu^x_{j+1/2}:~\mu^z_{j+1/2}\rightarrow(-1)^j\mu^z_{j+1/2};~~~~~\\
&\tilde{T_x}=T_x\cdot\mathcal{K}:~\mu^z_{j+1/2}\rightarrow\mu^z_{j+3/2};\\
&\phs=(\prod_j\rho^z_j)\cdot\mathcal{K}:~\mu^z_{j+1/2}\rightarrow\mu^z_{j+1/2};\\
&\mathcal{I}=(\prod_j\rho^z_j)\cdot\mathcal{O}_I:~~\mu^z_{j+1/2}\rightarrow\mu^z_{-j-1/2}.~~~\label{sym:spin:inversion:dual}
\eea
Most importantly, the Ising/parity symmetry $\mathcal{P}_f$ and inversion $\mathcal{I}$ anticommutes on the domain wall variable $\mu^z_{j+1/2}$ even on a periodic spin chain of an even length:
\bea
\mathcal{P}_f\cdot\mathcal{I}\circ\mu^z_{j+1/2}=-\mathcal{I}\cdot\mathcal{P}_f\circ\mu^z_{j+1/2}.
\eea
This projective symmetry action on the domain wall variable is captured by a nontrivial projective symmetry group~\cite{PhysRevB.65.165113}, i.e. a nontrivial group cohomology $\omega\in\mathcal{H}^2[G,Z_2]$ where $G$ is the symmetry group of the spin chain~\cite{CHEN20173}. This means destruction of the $AFM_z$ phase by condensing domain wall $\mu^z_{j+1/2}$ will inevitably breaks the symmetry, leading to e.g. an $AFM_x$ phase which spontaneously breaks the Ising symmetry. This is in parallel with 2D DQCPs where defects of one ordered phase carry nontrivial quantum numbers of another symmetry, and condensation of this defect will necessarily break another symmetry while restoring the originally broken symmetry.

Regarding the critical exponents of this critical point, $\nu$ can be related to the scaling dimension of the relevant perturbation $\cos(2\theta)$ by
\begin{equation}
\nu=\frac{1}{\mathfrak{d}-g}
\label{eq:llt_nu}
\end{equation}
in this TLL theory~\cite{PhysRevB.99.075103}.
Here $\mathfrak{d}=2$ is the space-time dimension. With the value of $\beta/\nu$ at hand and the general scaling relation $2\beta=\nu(\mathfrak{d}-2+\eta)$~\cite{francesco1997conformal}, we can immediately obtain the critical exponent $\eta$, namely the \emph{anomalous dimension} for the two point correlation function $\Gamma(n)=|n|^{2-\mathfrak{d}-\eta}$.
The scaling dimension for the order parameter reads $\text{dim}[M]=\eta/2=\beta/\nu$.

On one hand, from TABLE~\ref{tab:critical_exponents_d05d15} fixed with $\Delta^{\prime}=1.5$ we can see that, only for $\Delta=1.5$, these two order parameters vanish at the same critical point.
Numerical results of the critical exponents appears to be the same within numerical errors.
Moreover, they are consistent with Eq.~(\ref{eq:llt_nu}) and  (\ref{eq:llt_sd}), implying that it is indeed described by a $c=1$ TLL theory in the long-wavelength limit.
More data in TABLE~\ref{tab:critical_exponents_d15} suggests that the Luttinger parameter $g$ can vary within a finite range if $g<2$.
The scaling dimensions of $M_{x,z}$ are also the same at the critical point.
The emergent larger $U(1)_\theta$ symmetry unifies these two order parameters together and can rotate from one to the other.

On the other hand, if $\Delta=0.5$, the critical points for these two order parameters split, leading to a stable gapless phase between the two ordered phases.
Numerical results of the critical exponents contradict Eq.~(\ref{eq:llt_nu}) and  (\ref{eq:llt_sd}), which implies that this TLL theory is not valid any longer.
The analysis given by Abelian bosonization in Eq.~(\ref{eq:llt_sd}) seems to break down for a small $\Delta/t$.
However, interestingly, notice that the scaling dimensions of $M_{x,z}$ for the two phase boundaries where $M_{x,z}$ vanish respectively are still identical within numerical error, but become smaller than their values at the DQCP. Currently we do not have a good theoretical understanding of this gapless phase or how it emerges from the TLL at larger $\Delta/t$, and we leave these interesting questions for future works.

\begin{table}[hbt!]
    \caption{\label{tab:critical_exponents_d05d15}Critical point(s) and critical exponents for $\Delta^{\prime}=1.5$.}
\begin{ruledtabular}
\begin{tabular}{cccccc}
    $\Delta$ & $\left[V_{c}^{z}, V_{c}^{x}\right]$ & $\nu_{z}$ & $\beta_{z}/\nu_{z}$ & $\nu_{x}$ & $\beta_{x}/\nu_{x}$ \\
    \hline
    0.5 & [3.34, 2.88] & 1.47(1) & 0.19(7) & 1.51(3) & 0.20(6) \\
        1.0 & [4.10, 4.03] & 1.20(1) & 0.22(3) & 1.21(1) & 0.22(8) \\
    1.5 & [5.04, 5.04] & 0.99(7) & 0.23(5) & 1.00(3) & 0.24(2) \\
\end{tabular}
\end{ruledtabular}
\end{table}

\subsection{Finite-size analysis of the central charge}

As discussed in Appendix~\ref{sec:appdx_A} in detail, in a 1d chain of an odd length, there will be two exactly degenerate ground states with opposite fermion parities, due to the anti-commutation relation between inversion and parity symmetry operations. This leads an underestimated to entanglement entropy, and hence an underestimated central charge (see e.g. FIG.~\ref{fig:flsm_ee_lattice121120_d10dp15}(a)) by fitting the Cardy's formula Eq.~(\ref{eq:cardy_formula}) numerically.  

\begin{figure}[hbt!]
    \centering
    \includegraphics[width=0.49\textwidth]{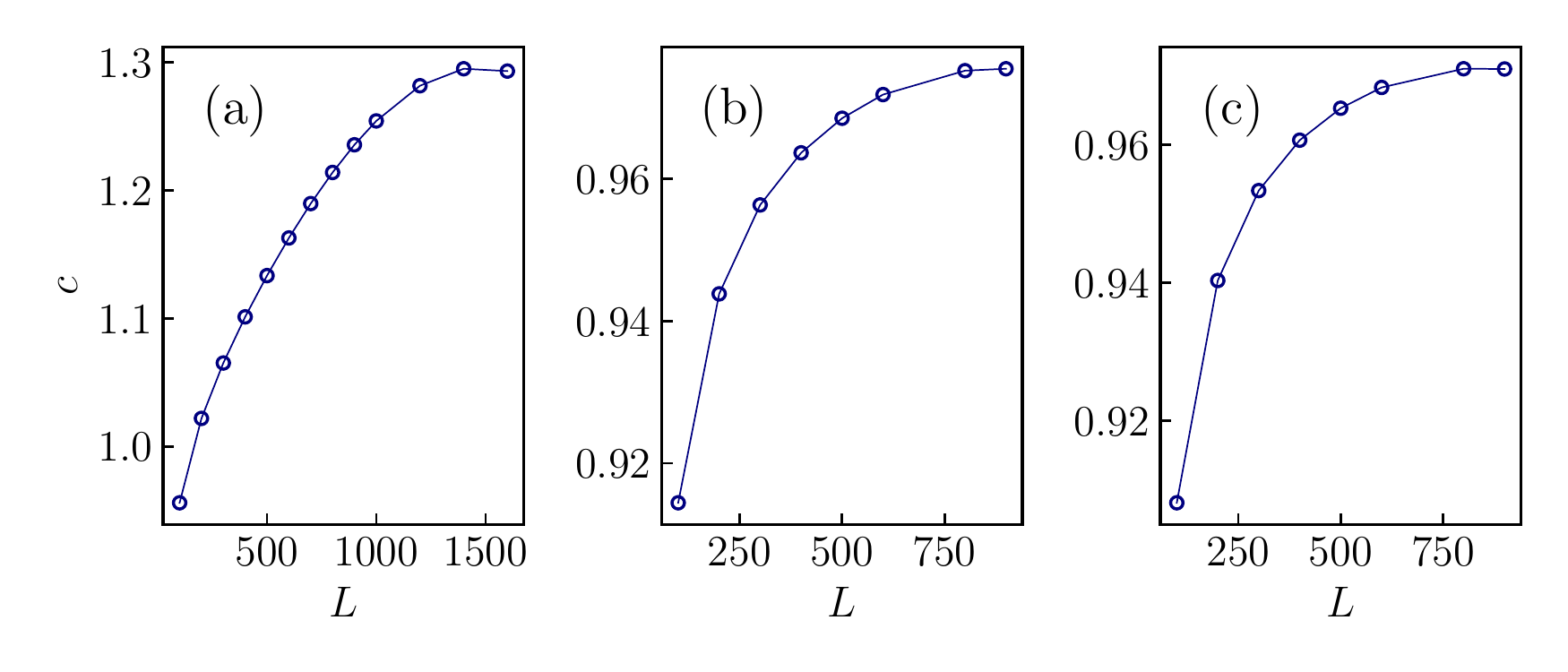}
    \caption{Finite-size scaling of fitted central charge $c$. $\Delta^{\prime}=1.5$ on even lattices. $\chi=256$. (a) $\Delta=0.5, V=3.00$. (b) $\Delta=1.5, V=5.04$. (c) $\Delta=1.0, V=-3.0$.}
    \label{fig:cc_scaling_even}
\end{figure}

To resolve this issue, we use an even system size, which splits the exact degeneracy in the spectrum, as illustrated in Appendix~\ref{sec:appdx_D}.
Even lattices allow us to extract the entanglement entropy and the central charge more reliably. Here we consider the same parameter range as in Sec.~\ref{sec:realize_DQCP}.
The results are shown in FIG.~\ref{fig:cc_scaling_even}.
At the DQCP in FIG.~\ref{fig:cc_scaling_even}(b), $\Delta=\Delta^\prime=1.5$, the critical exponents are given by $\nu=1$ and $g=1$ in TABLE~\ref{tab:critical_exponents_d05d15}. The corresponding central charge is approaching unity as the system size increases, consistent with a Tomonaga-Luttinger liquid (TLL). After the single DQCP splits into two phase boundaries, within the stable gapless phase, the central charge $c\rightarrow1.3$ as we compute up to $L=1600$ in FIG.~\ref{fig:cc_scaling_even}(a). This suggests that the stable gapless phase sandwiched by the two gapped symmetry-breaking phases cannot be described by a TLL with $c=1$. 

In FIG.~\ref{fig:cc_scaling_even}(c), we make the scaling analysis deep in the gapless phase at a negative $V=-3.0$. This stable gapless phase lies between the FM$_z$ phase and the Kitaev chain features a central charge of $c\rightarrow1$, again pointing to a TLL. 

\section{Concluding remarks \label{sec:conlusion}}

In this paper, we studied the phase diagram and quantum phase transitions in a 1d interacting fermion model with a Lieb-Schultz-Mattis (LSM) type anomaly. In the presence of a site-centered inversion symmetry, any gapped symmetric ground state must be a Kitaev chain with a Majorana zero mode on each open end. Via the Jordan-Wigner transformation, it is equivalent to a spin-$1/2$ model whose gapped ground states must break either the inversion or an Ising symmetry, which corresponds to the fermion parity in the fermion model. Such a LSM system provides a rich playground to identify unconventional quantum phase transitions between different ordered phases, not related to each other by spontaneous symmetry breakings, hence beyond the Ginzburg-Landau-Wilson paradigm. 

To understand the phase diagram of a generic fermion model with symmetric nearest-neighbor couplings, we first solve the non-interacting limit $V=0$. In the interacting cases with $V\neq{0}$, we implement the variational MPS method to numerically study the model. The phase diagram and phase boundaries are obtained using variance, entanglement entropy and ground state energy. Furthermore, we focus on the unconventional quantum phase transition between the inversion-breaking phase and the parity-breaking phase (i.e. the Kitaev chain), and carry out a detailed finite-size scaling analysis to extract critical exponents. This is combined with Abelian bosonization and projective symmetry group analysis, to understand the nature of this ``deconfined''quantum critical point (DQCP), where both inversion-breaking and parity-breaking order parameters vanish simultaneously. We find that the numerically measured critical exponents are captured by the Luttinger parameter in a $U(1)$ Luttinger liquid. We have also identified a stable gapless phase, which is symmetric and separates the two symmetry-breaking phases. While the nature of this gapless phase and how it emerges from the DQCP remains unknown, we leave this interesting question for future works.

\acknowledgments

WZ acknowledges helpful discussions with Xiao-Qi Sun, Shuo Yang and Shuai A. Chen. This work is supported by National Science Foundation under award number NSF DMR-1653769 (WZ,YML) and  Partnerships for Research and Education in Materials Grant No. NSF DMR-1828019 (DNS).

\appendix

\section{Symmetry implementations on the fermion chain vs. the spin chain}\label{app:symmetry}

Here we address in detail the symmetry operations in the fermion language vs. the spin language.

In the fermion model,
\begin{equation}
    \begin{aligned}
        H
        &=\sum_{j}(-)^{j}\left[t\left(c_{j}^{\dagger}c_{j+1}+h.c.\right)+\left(\Delta c_{j}^{\dagger}c_{j+1}^{\dagger}+h.c.\right)\right] \\
        &+\left(\text{i}\Delta^{\prime}c_{j}^{\dagger}c_{j+1}^{\dagger}+h.c.\right)+V\left(n_{j}-\frac{1}{2}\right)\left(n_{j+1}-\frac{1}{2}\right)
    \end{aligned}
    \label{model:fermion}
\end{equation}
By the Jordan-Wigner transformation in Eq.~(\ref{eq:jordan-wigner}),
the above fermion model is transformed into a spin chain
\begin{equation}
H_\text{spin}=\sum_j\sum_{\alpha=x,y,z}J_\alpha S_j^\alpha S_{j+1}^\alpha+(-)^j\Gamma\left(S_j^xS_{j+1}^y+S_j^yS^x_{j+1}\right)
\label{model:spin}
\end{equation}
where the exchange couplings are given by
\bea
&\notag J_x=2(t+\Delta),~~J_y=2(t-\Delta),~~J_z=V,\\
&\Gamma=-2\Delta^\prime.
\eea
Three symmetries are clearly present in the spin model:
\bea
&\mathcal{P}_f=\prod_jZ_j:~(S_j^x,S_j^y,S_j^z)\rightarrow(-S_j^x,-S_{j}^y,S^z_{j}),~~~\label{eq:sym:Ising}\\
&\phs=(\prod_jX_j)\cdot\mathcal{K}:~(S_j^x,S_j^y,S_j^z)\rightarrow(S_j^x,S_{j}^y,-S^z_{j}),~~~\label{eq:sym:phs}\\
\label{eq:sym:magnetic translation}
&\tilde T_x=T_x\cdot\mathcal{K}:~(S_j^x,S_j^y,S_j^z)\rightarrow(S_{j+1}^x,-S_{j+1}^y,S^z_{j+1}).~~
\eea
They are nothing but the magnetic translation (\ref{eq:magnetic translation}), fermion parity (i.e. Ising symmetry) (\ref{sym:spin:ising}) and anti-unitary particle-hole symmetry (\ref{eq:particle-hole sym}) discussed in the fermion context.

Although the Hamiltonian remains local in both the fermion and the spin representations, due to the Jordan-Wigner string, a locality-preserving symmetry operation in one representation may appear to be non-local in the other representation. One example is the inversion symmetry (\ref{eq:inversion}) discussed in this work. Below we write down two possible sets of inversion symmetry operations: the first one is non-local in the spin language; the second one preserves locality in the spin language but looks non-local in the fermion language. Since all numerical simulations are carried out in the spin representation, in the main text we will stick to the 2nd set of inversion symmetry summarized in Appendix \ref{app:local sym}.

\subsection{Non-local inversion symmetry in the spin representation}\label{app:non-local sym}

We first consider the following inversion symmetry
\bea
c_j\overset{\mathcal{I}}\longrightarrow\imth c_{-j}^\dagger
\eea
in the fermion chain. On an open spin chain of length $L=2N+1$, the associated inversion symmetry generator in the spin model (\ref{eq:spin_ham 1}) is
\bea\label{eq:non local inversion}
\mathcal{I}=e^{\imth\frac\pi4\sum_j(-1)^{j+N+1}\sigma^z_j}\cdot e^{\imth\frac\pi4\mathcal{P}_f}(\prod_r\sigma^x_r)\cdot\mathcal{O}_I.
\eea
where $\mathcal{O}_I$ is the spatial inversion operator.

In terms of the spin language, the parity operator $\mathcal{P}_{f}=\prod_{l=-N}^{N}\left(1-2c_{l}^{\dagger}c_{l}\right)=\prod_{l=-N}^{N}\left(-\sigma_{l}^{z}\right)$.
Note that $\sigma_{j}^{+}\sigma_{j}^{z}=-\sigma_{j}^{+}, \sigma_{j}^{z}\sigma_{j}^{+}=\sigma_{j}^{+}$.
Thus we have $\left\{\mathcal{P}_{f}, \sigma_{j}^{+}\right\}=0$.
Similarly, $\left\{\mathcal{P}_{f}, \sigma_{j}^{-}\right\}=0$.
Since we have the inverse Jordan-Wigner transformation $\sigma_{j}^{z}=2c_{j}^{\dagger}c_{j}-1, \sigma_{j}^{-}=\prod_{l=-N}^{j-1}\left(-\sigma_{l}^{z}\right)c_{j}, \sigma_{j}^{+}=c_{j}^{\dagger}\prod_{l=-N}^{j-1}\left(-\sigma_{l}^{z}\right)$.
Thus we can find that, under the inversion symmetry $\mathcal{I}$,
\begin{equation}
    \begin{aligned}
        \sigma_{j}^{z}&\rightarrow -\sigma_{-j}^{z}, \\
        \sigma_{j}^{-}&\rightarrow \text{i}(-)^{N+j}\sigma_{-j}^{+}\mathcal{P}_{f}=\text{i}(-)^{N+j+1}\mathcal{P}_{f}\sigma_{-j}^{+}, \\
        \sigma_{j}^{+}&\rightarrow \text{i}(-)^{N+j+1}\mathcal{P}_{f}\sigma_{-j}^{-}.
    \end{aligned}
    \label{eq:spin_sym_trans}
\end{equation}

Therefore, due to the Jordan-Wigner string, the above inversion symmetry $\mathcal{I}$ is not a locality-preserving unitary:
\bea
\notag&S_j^x\overset{\mathcal{I}}\longrightarrow(-1)^{j+N+1}\mathcal{P}_f\cdot S_{-j}^y,\\
\notag&S_j^y\overset{\mathcal{I}}\longrightarrow(-1)^{j+N+1}\mathcal{P}_f\cdot S_{-j}^x,\\
&S_j^z\overset{\mathcal{I}}\longrightarrow-S^z_{-j}.
\eea
Interestingly, a (non-local) string order parameter is required to preserve this non-local inversion symmetry, unlike the usual local order parameter for the case of a locality-preserving symmetry. This is discussed in more detail later, in Appendix \ref{app:string order}.

\subsection{Locality-preserving inversion symmetry in the spin representation}\label{app:local sym}

Alternatively, there is also a locality-preserving inversion symmetry preserved in the spin chain (\ref{model:spin}):
\bea\label{eq:local inversion}
\mathcal{I}=(\prod_jX_j)\cdot\mathcal{O}_I
\eea
under which the spins transform as
\bea
(S_j^x,S_j^y,S_j^z)\overset{\mathcal{I}}\longrightarrow(S_{-j}^x,-S_{-j}^y,-S_{-j}^z).
\eea
This symmetry, however does not have a local form in the fermion language:
\bea
c_j\overset{\mathcal{I}}\longrightarrow-\mathcal{P}_fc_{-j}^\dagger
\eea
We shall stick to this locality-preserving inversion symmetry in the main text.

\section{Zero-mode, entanglement entropy and finite-size analysis in the XY model on even and odd open chains \label{sec:appdx_A}}

In this section, we use the pedagogical 1D XY model to illustrate the issue of zero-mode and its effects on the EE on even and odd lattices under OBC, which is similar to our model when it comes to the lower EE and underestimation of the central charge.
Different DMRG methods could also make some subtle difference.
The Hamiltonian is
\begin{equation}
    H_{XY}=
    \sum_{j=0}^{L-2}\left[\left(\frac{1+\gamma}{2}\right)\sigma_{j}^{x}\sigma_{j+1}^{x}+\left(\frac{1-\gamma}{2}\right)\sigma_{j}^{y}\sigma_{j+1}^{y}\right],
    \label{eq:xy_ham}
\end{equation}
where $\gamma$ is a free parameter.
$\gamma=0$ is the critical point, at which the system becomes gapless.
Eq.~(\ref{eq:xy_ham}) is equivalent to the free 1D superconductor model
\begin{equation}
   H_{f\text{-}XY}
   =\sum_{j=0}^{L-2}\left[\left(c_{j}^{\dagger}c_{j+1}-c_{j}^{\dagger}c_{j+1}\right)+\gamma\left(c_{j}^{\dagger}c_{j+1}^{\dagger}-c_{j}^{\dagger}c_{j+1}^{\dagger}\right)\right].
   \label{eq:xy_ham_f}
\end{equation}
By a unitary transformation, the single quasi-particle spectrum of Eq.~(\ref{eq:xy_ham_f}) can be computed exactly~\cite{vanHemmen1980} as $H_{f\text{-}XY}=\sum_{k=0}^{L-1}\lambda_{k}\left(d_{k}^{\dagger}d_{k}-\frac{1}{2}\right)$.
The many-body excitation spectrum is given by various filling combinations of the single quasi-particle spectrum.
For $\gamma=0$, on a finite odd lattice $L$, we find that there is always an \emph{exact} zero-mode $\lambda_{0}=0$.
While on a even lattice, there is a  finite-size non-zero but very small gap $\lambda_{0}\neq{0}$.
That is, on a finite odd lattice with OBC, the two ground states $|\Psi_{0}\rangle=|0\rangle$ and $|\Psi_{1}\rangle=d_{0}^{\dagger}|0\rangle$ are precisely degenerated.
They belong to different topological sectors characterized by the fermion parity $\mathcal{P}_{f}$.

\begin{figure}[hbt!]
     \centering
     \includegraphics[width=0.49\textwidth]{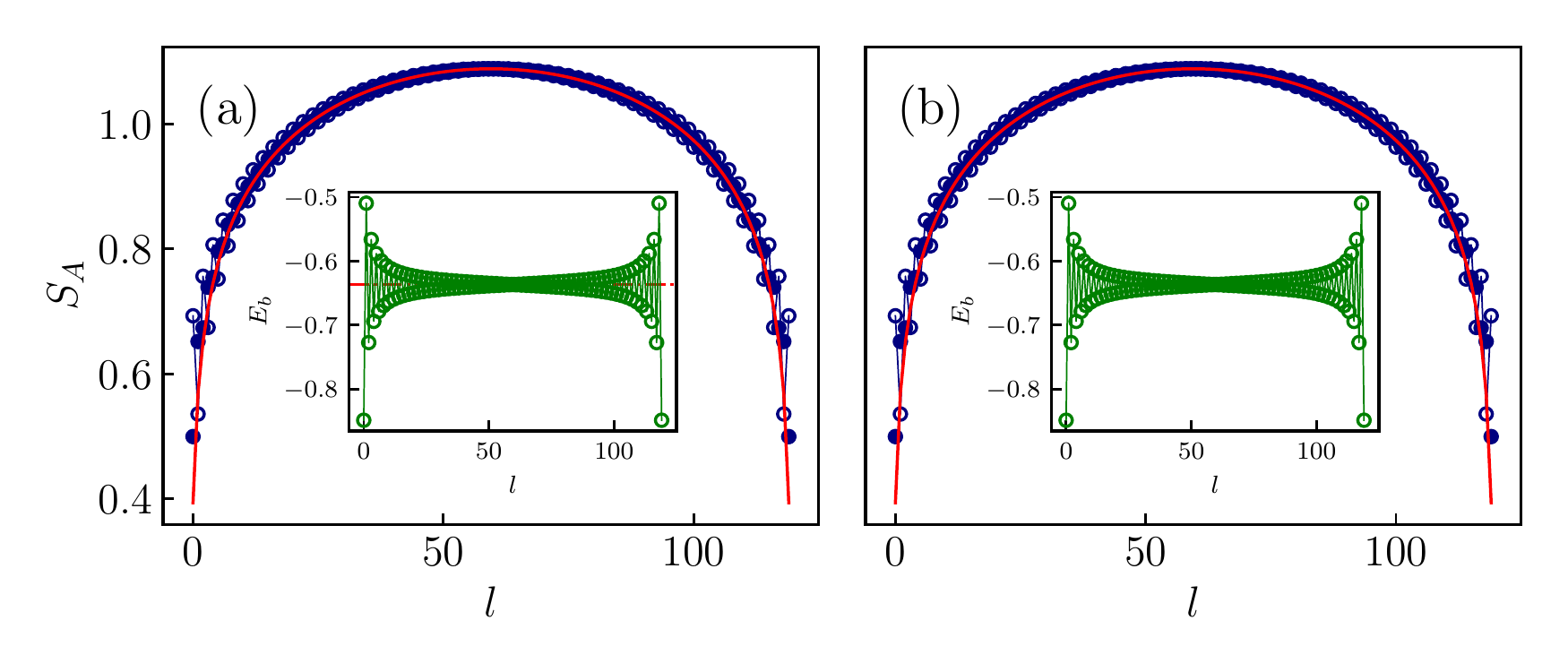}
     \caption{Bond energy $E_{b}$ (insets) and bipartite EE $S_{A}$ in the XY spin chain under OBC computed from the exact solution. $L=121$. Hollow circles represent the original data while filled circles represent the extracted uniform EE as defined in Eq.~(\ref{eq:entanglement_entropy_parts}).  (a) $|\Psi_{0}\rangle$ without zero-mode. (b) $|\Psi_{1}\rangle$ with zero-mode. Both fitted central charges are $c\approx{0.962}$.}
     \label{fig:xychain_ee_lattice121_two_gs}
\end{figure}

\begin{figure}[hbt!]
     \centering
     \includegraphics[width=0.49\textwidth]{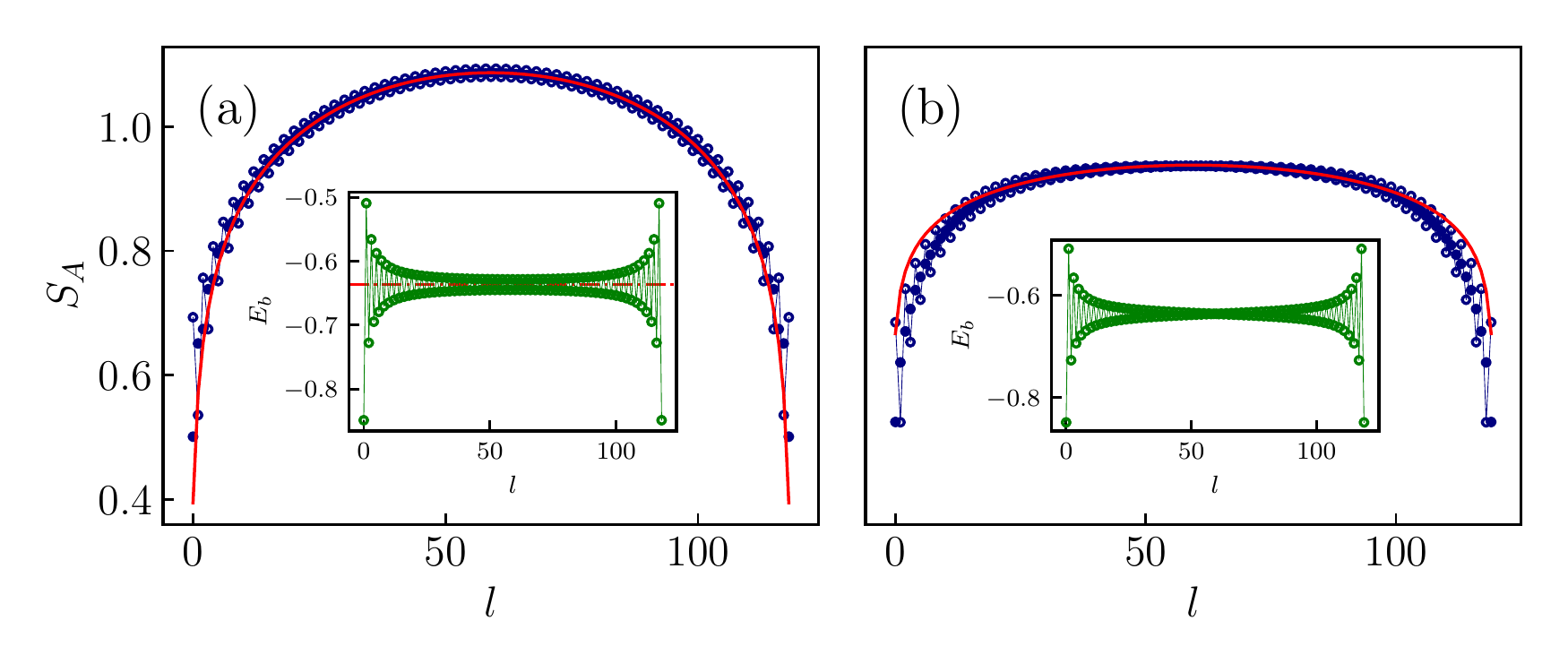}
     \caption{Bond energy $E_{b}$ (insets) and bipartite EE $S_{A}$ in the open XY chain obtained by MPS with $\chi=64$. Hollow circles represent the original data while filled circles represent the extracted uniform EE as defined in Eq.~(\ref{eq:entanglement_entropy_parts}).  (a) $L=120$. Fitted central charge $c\approx{0.960}$. (b) $L=121$. Fitted central charge $c\approx{0.375}$.}
     \label{fig:xychain_ee_lattice120121}
\end{figure}

If the ground state of Eq.~(\ref{eq:xy_ham_f}) is a Slater determinant, the reduced density matrix of a subsystem $A$ containing $M$ sites can be written as $\rho_{A}=e^{-\mathcal{H}_{A}}/\mathcal{Z}$ and its bipartite entanglement spectrum $\{\omega\}$ can be analytically extracted from the correlation matrix~\cite{Peschel_2003, Peschel_2009, Calabrese_2016}.
$\mathcal{Z}$ is the partition function.
Therefore the corresponding EE reads
\begin{equation}
    \begin{aligned}
        S_{A}
        &=-\text{tr}\left( \rho_{A}\ln\rho_{A} \right) \\
        &=-\sum_{l=0}^{M-1}\left( \frac{\omega_{l}}{2} \right)\tanh\left( \frac{\omega_{l}}{2} \right)+\sum_{l=0}^{M-1}\ln\left[ 2\cosh\left( \frac{\omega_{l}}{2} \right) \right].
    \end{aligned}
    \label{eq:}
\end{equation}
We find that bond energy and EE in $|\Psi_{0}\rangle$ and $|\Psi_{1}\rangle$ are the same as shown in FIG.~\ref{fig:xychain_ee_lattice121_two_gs}.
However, EE in the superposed state $|\Psi\rangle=\alpha|\Psi_{0}\rangle+\beta|\Psi_{1}\rangle$ cannot be analytically computed since the superposition of two Slater determinants may not be written as another Slater determinant.
If we simulate the XY chain using a randomly initialized MPS, we can converge to the minimally entangled state, which turns out to the superposition of $|\Psi_{0, 1}\rangle$ and results in a lower EE as well as a underestimation of the central charge.
They are illustrated in FIG.~\ref{fig:xychain_ee_lattice120121} and we think this is the reason for the incommensurability observed in other odd spin chains~\cite{PhysRevB.87.094415}.
If we add boundary perturbations such as  $H_{1}=h\left(\sigma_{0}^{z}+\sigma_{L-1}^{z}\right)$ in the early sweeping stage to select the MPS within a fixed parity sector, numerically we indeed can obtain the results in FIG.~\ref{fig:xychain_ee_lattice121_two_gs} for odd lattices.

\begin{figure}[hbt!]
     \centering
     \includegraphics[width=0.49\textwidth]{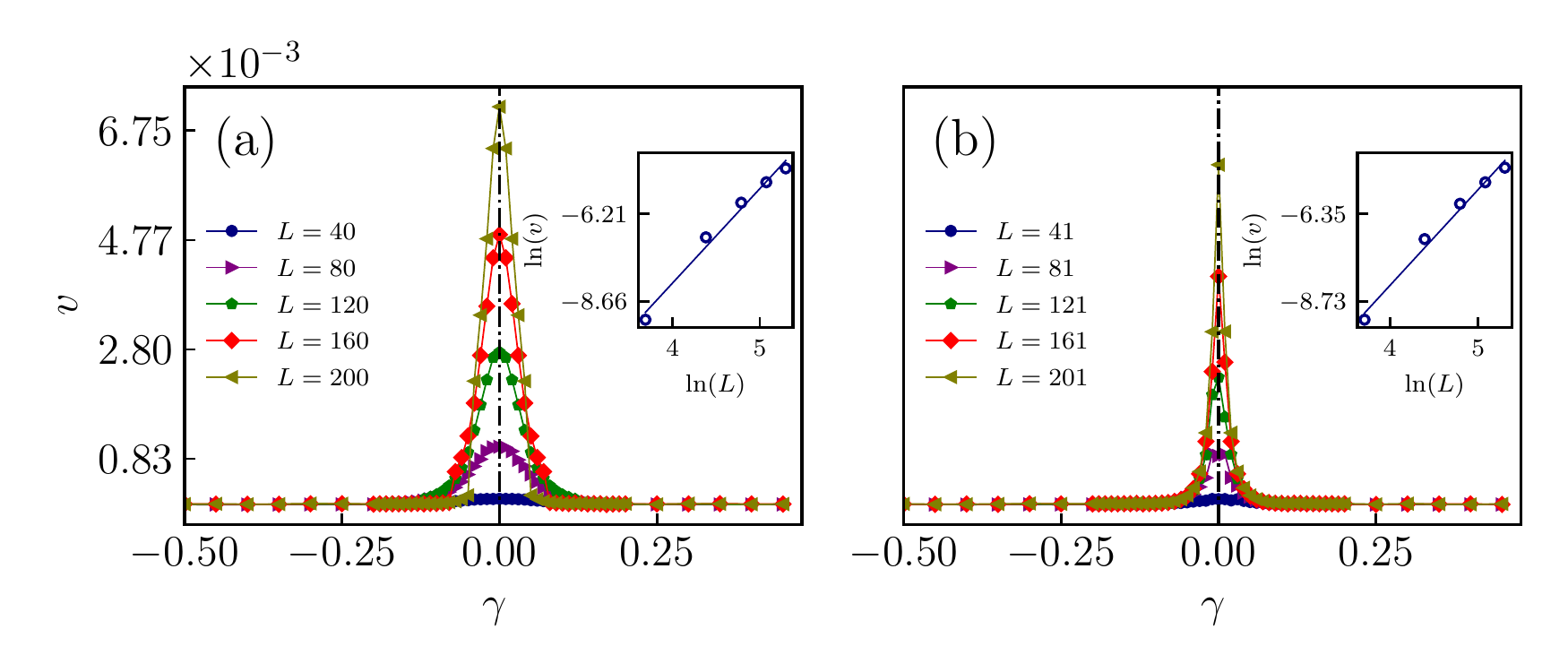}
     \caption{Variance of the open XY chain. $\chi=64$. Insets show the logarithm fittings of $v$ at the critical point. (a) Even lattices. (b) Odd lattices.}
     \label{fig:xychain_variance}
\end{figure}

\begin{figure}[hbt!]
     \centering
     \includegraphics[width=0.49\textwidth]{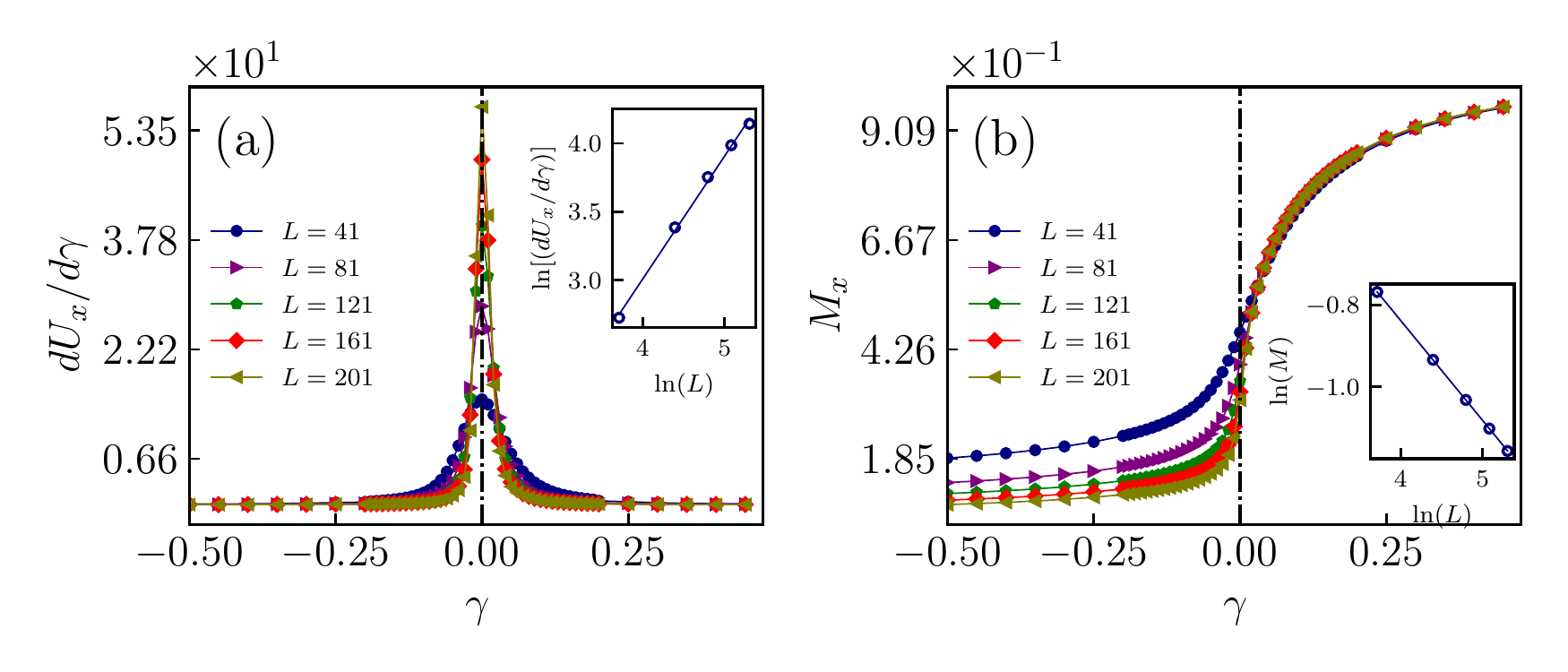}
     \caption{(a) Binder cumulant's derivative and (b) magnetization in XY model  on odd chains. $\chi=64$. Fitted critical exponents as given by insets: $\nu_{x}=1.11(1), \beta_{x}/\nu_{x}=0.24(4)$.}
     \label{fig:xychain_binder_mag_odd}
\end{figure}

\begin{figure}[hbt!]
     \centering
     \includegraphics[width=0.49\textwidth]{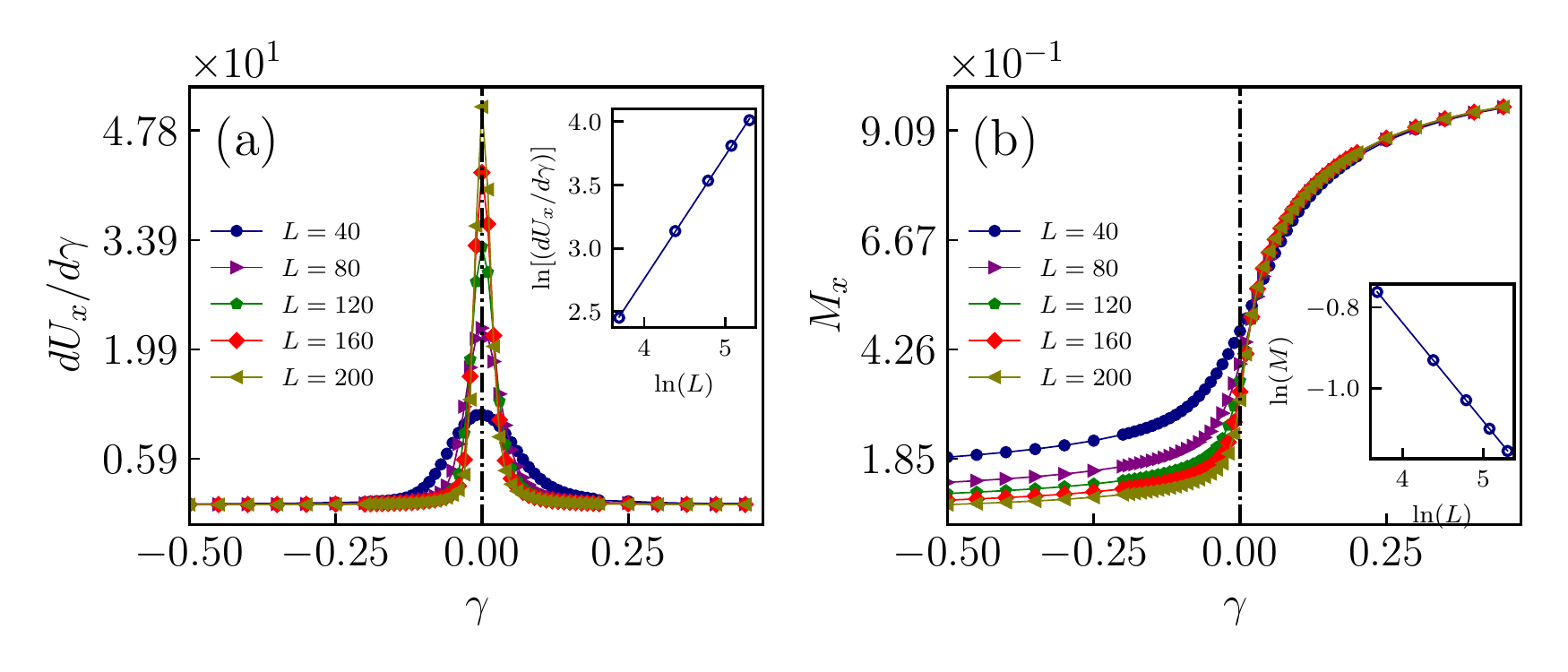}
     \caption{(a) Binder cumulant's derivative and (b) magnetization in XY model  on even chains. $\chi=64$. Fitted critical exponents as given by insets: $\nu_{x}=1.02(9), \beta_{x}/\nu_{x}=0.24(4)$.}
     \label{fig:xychain_binder_mag_even}
\end{figure}

Next we carry out some finite-size analysis in terms of even and odd lattices for the XY model.
From FIG.~\ref{fig:xychain_variance} we can see that the variance $v$ is sharper on odd lattices.
We consider the Neel order parameter
\begin{equation}
M_{x}
=\frac{1}{L}\sum_{j=0}^{L-1}(-)^{j}\sigma_{j}^{x}
\end{equation}
and the corresponding Binder culmulant $U_{x}$.
From FIG.~\ref{fig:xychain_binder_mag_odd} and FIG.~\ref{fig:xychain_binder_mag_even} we can see that on even and odd lattices, the XY model exhibits almost similar critical properties up to some numerical errors.
However, their EE can be dramatically different.

\section{Larger bond-dimension test \label{sec:appdx_B}}

\begin{figure}[hbt!]
     \centering
     \includegraphics[width=0.49\textwidth]{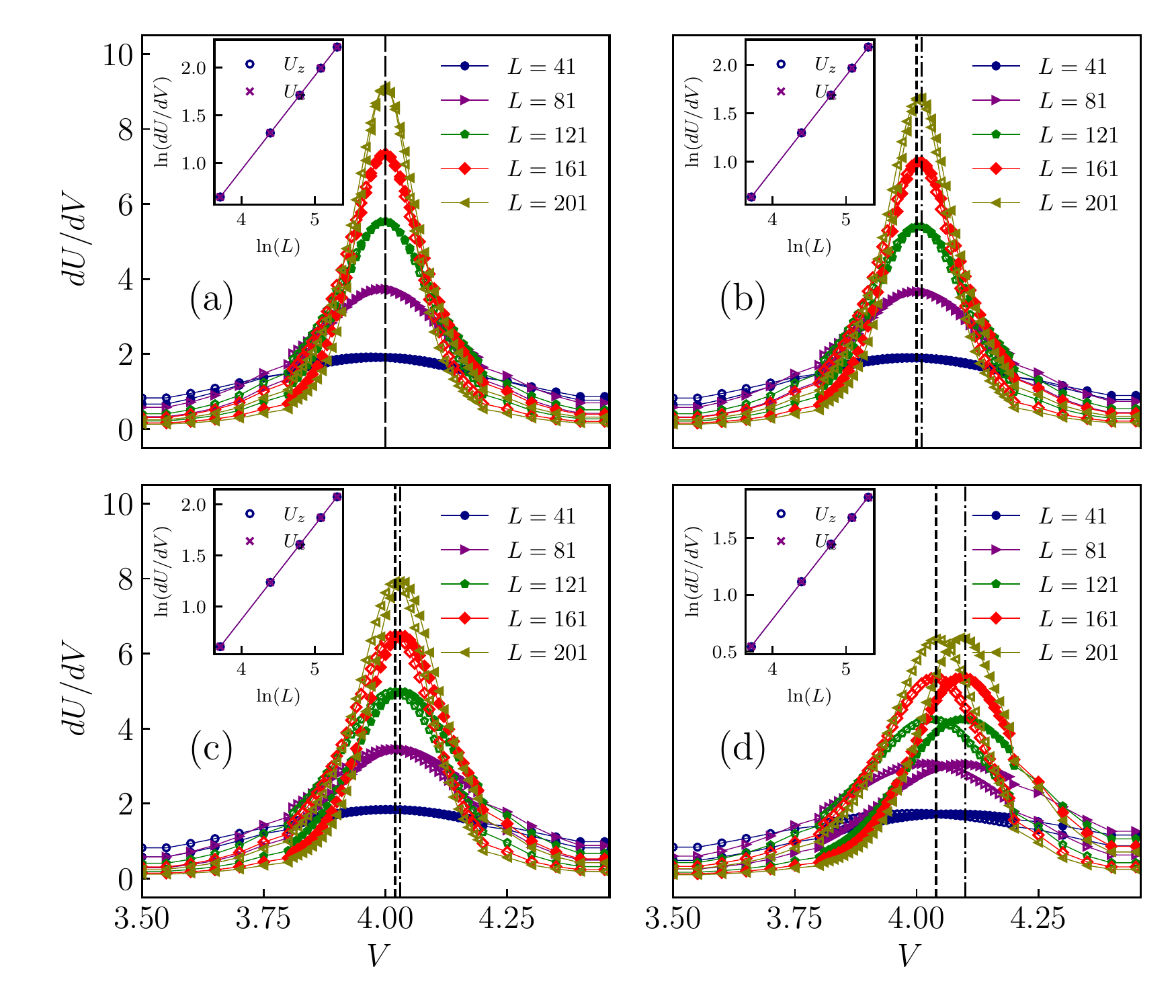}
     \caption{Derivatives of the Binder cumulants $dU_{z}/dV$ (solid markers) and $dU_{x}/dV$ (hollow markers) around the critical point. $\chi=128$. $\Delta=1.0$. (a, b, c, d) denote $\Delta^{\prime}=(0.0, 0.5, 1.0, 1.5)$, respectively.
     }
     \label{fig:flsm_binder_diff_odd_d10_chi128}
\end{figure}

\begin{figure}[hbt!]
    \centering
    \includegraphics[width=0.49\textwidth]{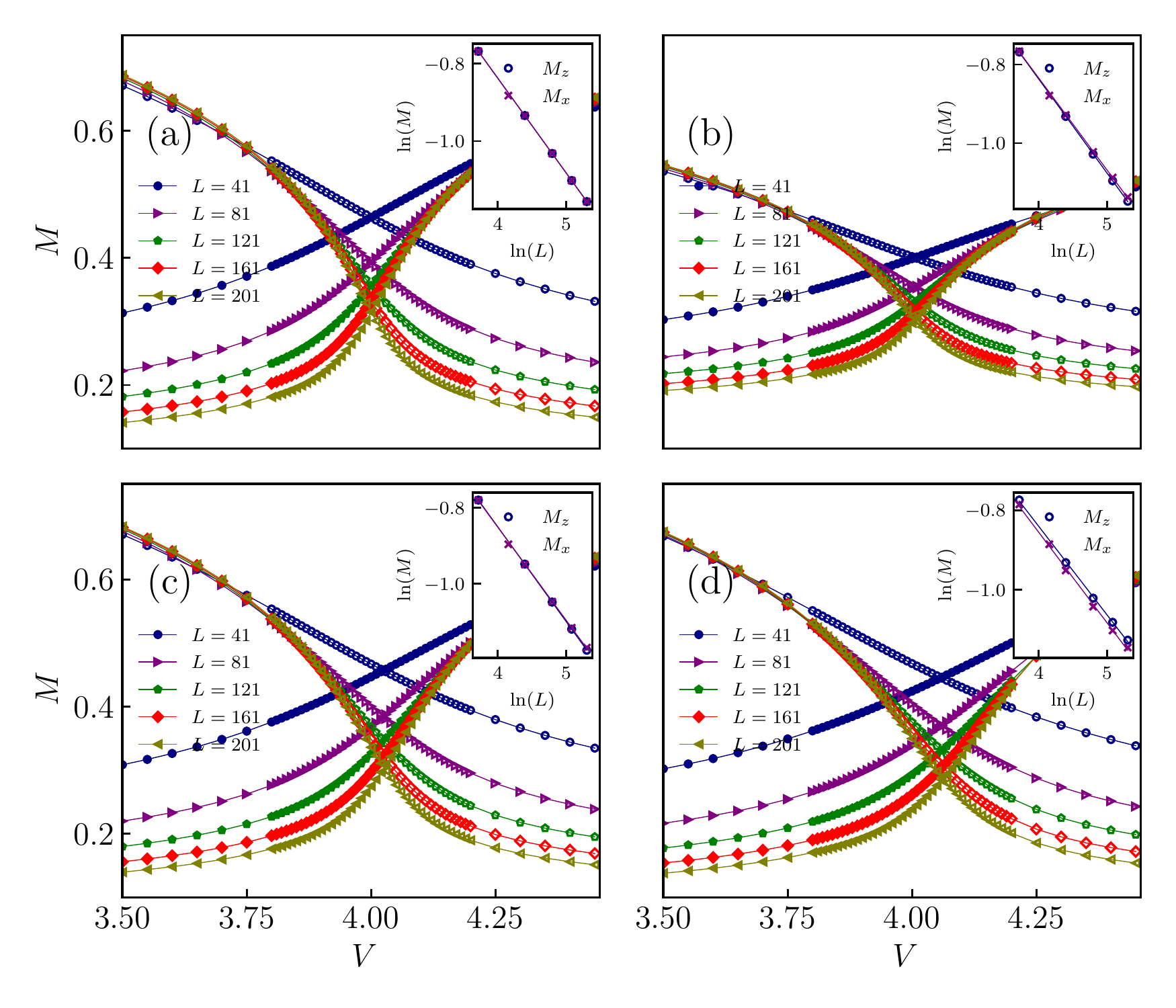}
    \caption{Magnetic order parameters $M_{\text{AFM}-z}$ (monotonically increasing solid data markers) and $M_{x}$ (monotonically decreasing hollow data markers) around the critical point. $\chi=128$. $\Delta=1.5$. (a, b, c, d) denote $\Delta^{\prime}=(0.0, 0.5, 1.0, 1.5)$, respectively.}
    \label{fig:flsm_mag_odd_d10_chi128}
\end{figure}

To test the convergence of our numerical computation, we also repeat the simulation with the same parameters as in FIG.~\ref{fig:flsm_binder_diff_odd_d10} and FIG.~\ref{fig:flsm_mag_odd_d10} up to the bond-dimension $\chi=128$.
They are re-plotted as FIG.~\ref{fig:flsm_binder_diff_odd_d10_chi128} and FIG.~\ref{fig:flsm_mag_odd_d10_chi128}.
The critical exponents obtained with $\chi=128$ are summarized in TABLE~\ref{tab:critical_exponents_d10_chi128}, in which we find the numbers are almost identical to those shown in TABLE~\ref{tab:critical_exponents_d10}.
In this sense, we claim that our numerical simulation has already well converged with $\chi=64$ and the error-bar estimation is faithful.

% chi=128
\begin{table}[hbt!]
    \caption{\label{tab:critical_exponents_d10_chi128}Critical point(s) and critical exponents for $\Delta=1.0$. $\chi=128$.}
\begin{ruledtabular}
\begin{tabular}{cccccc}
    $\Delta^{\prime}$ & $\left[V_{c}^{z}, V_{c}^{x}\right]$ & $\nu_{z}$ & $\beta_{z}/\nu_{z}$ & $\nu_{x}$ & $\beta_{x}/\nu_{x}$ \\
    \hline
    0.0 & [4.00, 4.00] & 1.01(2) & 0.24(4) & 1.01(2) & 0.24(4) \\
    0.5 & [4.01, 4.00] & 1.02(8) & 0.23(9) & 1.03(0) & 0.23(3) \\
    1.0 & [4.03, 4.02] & 1.08(5) & 0.24(9) & 1.08(5) & 0.24(5) \\
    1.5 & [4.10, 4.03] & 1.20(7) & 0.22(3) & 1.21(2) & 0.22(8) \\
\end{tabular}
\end{ruledtabular}
\end{table}

\section{A string order parameter that preserves non-local inversion symmetry (\ref{eq:non local inversion})}\label{app:string order}

We can construct another kind of non-local string order parameter according to the rule given by Eq.~(\ref{eq:spin_sym_trans}).
It is invariant under the inversion symmetry $\mathcal{I}$ and looks like
\begin{equation}
    \begin{aligned}
        &{O}_{\text{Str-}x}
        \equiv\sum_{j=-N}^{N}\frac{1+\text{i}(-)^{N+j+1}Q}{2}\sigma_{j}^{x} \\
        &=\sum_{j=-N}^{N}\frac{\sigma_{j}^{x}+(-)^{N+j+1}\sigma_{j}^{y}\prod_{l=-N, l\neq{j}}^{N}\sigma_{l}^{z}}{2}.
    \end{aligned}
    \label{eq:}
\end{equation}
To write ${O}_{\text{Str-}x}$ in a MPO form, we have to encode the operator string into the productions of matrices living on \emph{each} site.
It turns out that the corresponding MPO has a dimension of $D=4N+2$ and can be written as
\begin{equation}
    V_{\text{Str-}x}^{[j]}=\frac{1}{2}
    \begin{pmatrix}
        \mathbbm{1} & \dots & 0 \\
        \vdots & \ddots & \vdots \\
        \sigma^{x} & \dots & \mathbbm{1}
    \end{pmatrix}.
    \label{eq:}
\end{equation}
On different sites, $V_{\text{Str-}x}^{[j]}$ has different forms.
For $j=-N$, $\left(2V_{\text{Str-}x}^{[-N]}\right)_{D-1, D-2}=-\sigma^{y}; \left(2V_{\text{Str-}x}^{[-N]}\right)_{D-1, D-2-l}=\sigma^{z}, l=1, \dots, 2N$.
For $j=N$, $\left(2V_{\text{Str-}x}^{[N]}\right)_{1, 0}=-\sigma^{y}; \left(2V_{\text{Str-}x}^{[N]}\right)_{1+l, 0}=\sigma^{z}, l=1, \dots, 2N$.
For $-N<j<N$,
\begin{equation}
    \begin{aligned}
    &\left(2V_{\text{Str-}x}^{[j]}\right)_{D-1-(N+j)-l, D-2-(N+j)-l} \\
    &=\left\{
        \begin{aligned}
            &\sigma^{z},~ l\neq(N+j); \\
            &(-)^{N+j+1}\sigma^{y},~ l=(N+j)
        \end{aligned}
    \right.
    \end{aligned}
    \label{eq:}
\end{equation}
for $l=0, \dots, 2N$.

\begin{figure}[hbt!]
    \centering
    \includegraphics[width=0.49\textwidth]{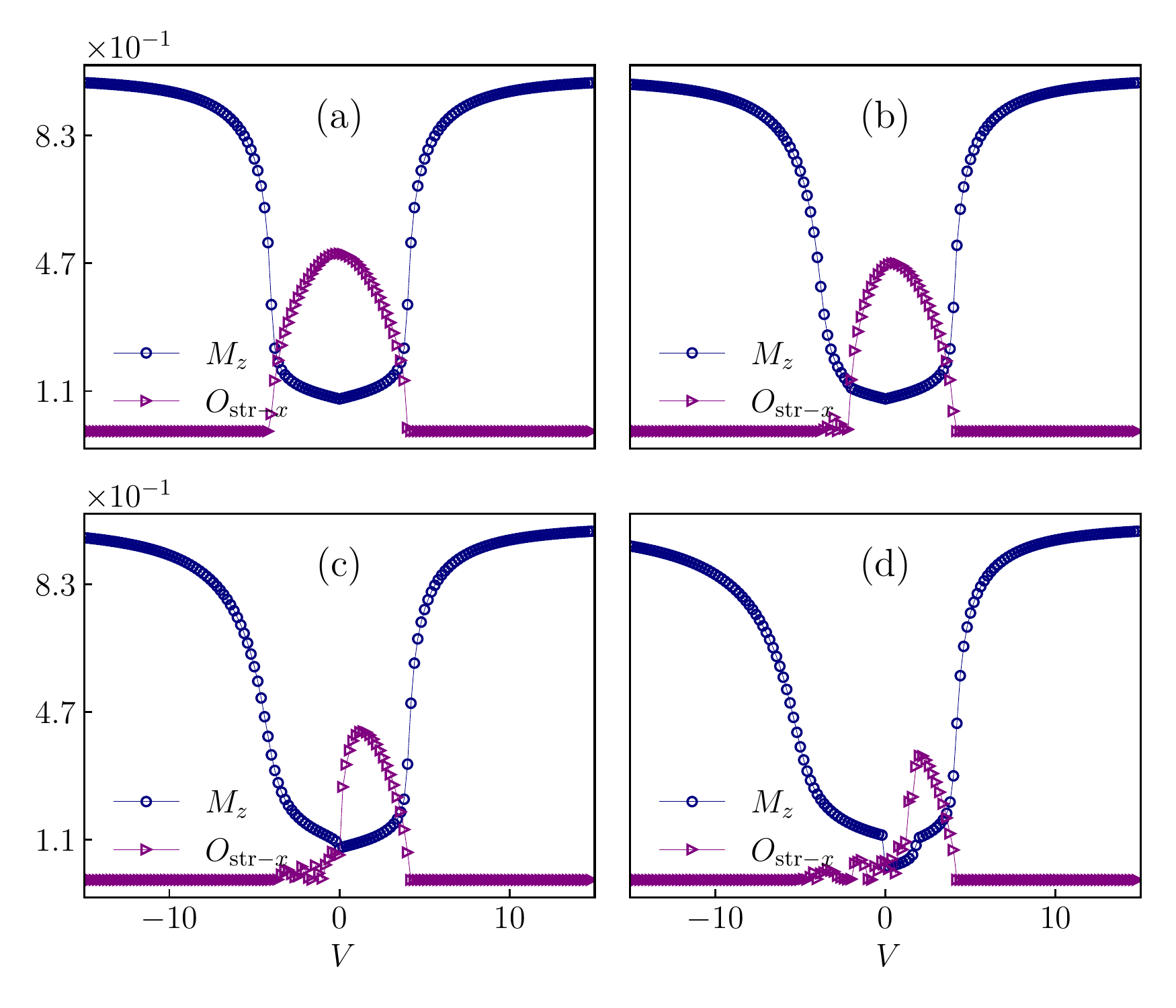}
    \caption{String order parameter and $z$-magnetic order parameter. $\Delta=1.0$. $L=121, \chi=64$. (a, b, c, d) denote $\Delta^{\prime}=(0.0, 0.5, 1.0, 1.5)$, respectively.}
    \label{fig:stringx}
\end{figure}

From FIG.~\ref{fig:stringx} we can see that in the gapped TSC phases, the non-vanishing string order parameter $\langle{O}_{\text{Str-}x}\rangle$ does imply the spontaneous symmetry breaking of the associated parity symmetry $\mathcal{P}_f$, which is consistent with the results given by the local order parameter $M_{x}$.

\section{Numerical results on  on even lattices as a perturbative analysis\label{sec:appdx_D}}

On a finite lattice under OBC, although the symmetry $\mathcal{I}$ is only precisely well-defined on odd lattices as we discussed in the main text, we can regard the even ones as a kind of perturbation on the boundary by removing one site.

\subsection{Critical exponents at the critical point(s)}

In TABLE~\ref{tab:critical_exponents_d10_even} we perform the finite-size analysis on even lattices following the same parameters in TABLE~\ref{tab:critical_exponents_d10}, which show that although specific numbers are different but they are close and follow the same trend.

\begin{table}[H]
    \caption{\label{tab:critical_exponents_d10_even}Critical exponents at $\Delta=1.0$ on even lattices}
\begin{ruledtabular}
\begin{tabular}{cccccc}
    $\Delta^{\prime}$ & $\left[V_{c}^{z}, V_{c}^{x}\right]$ & $\nu_{z}$ & $\beta_{z}/\nu_{z}$ & $\nu_{x}$ & $\beta_{x}/\nu_{x}$ \\
    \hline
    0.0 & [4.00, 4.00] & 1.00(7) & 0.24(4) & 1.00(7) & 0.24(4) \\
    0.5 & [4.01, 4.00] & 1.02(3) & 0.24(1) & 1.02(3) & 0.23(7) \\
    1.0 & [4.04, 4.01] & 1.06(8) & 0.23(6) & 1.07(1) & 0.23(2) \\
    1.5 & [4.13, 4.01] & 1.14(9) & 0.19(3) & 1.16(3) & 0.20(9) \\
\end{tabular}
\end{ruledtabular}
\end{table}

\subsection{Central charge in the gapless phase}

\begin{figure}[hbt!]
    \centering
    \includegraphics[width=0.49\textwidth]{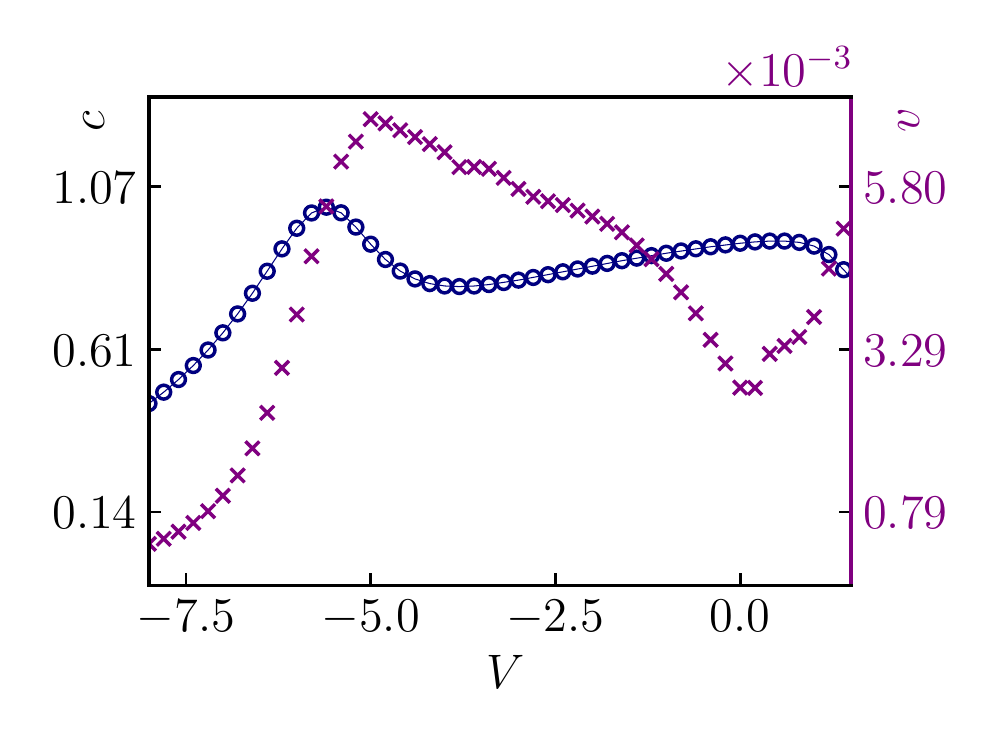}
    \caption{Fitted central charge $c$ and variance $v$ in the negative-$V$ gapless phase for $\Delta=1.0$ and $\Delta^{\prime}=1.5$ on an even lattice $L=120$. $\chi=64$.}
    \label{fig:flsm_cc_var_lattice120_chi64_d10dp15}
\end{figure}

Away from the phase boundaries namely deeply in the gapless phase induced by non-vanishing $\Delta^{\prime}$, we believe that Cardy's formula could work well.
In FIG.~\ref{fig:flsm_cc_var_lattice120_chi64_d10dp15} we show that the fitted central charge as well as the variance on an even lattice $L=120$.
Furthermore, we also present several representative examples using larger lattices $L=400$ and $L=800$ with the bond dimension $\chi=256$ to make more accurate estimation of the central charge as shown in TAB.~\ref{tab:central_charge_gapless}.

\begin{table}[H]
    \caption{\label{tab:central_charge_gapless} Fitted central charge $c$ in the negative-$V$ gapless phase for $\Delta=1.0, \Delta^{\prime}=1.5$ on other larger even lattices. $\chi=256$ for $L=400, 800$. $\chi=512$ for $L=1000$.}
\begin{ruledtabular}
\begin{tabular}{ccccccc}
    $V$ & -4.0 & -3.0 & -2.0 & -1.0 & 0.0 & 1.0 \\
    \hline
    $L=400$ & 0.949 & 0.961 & 0.982 & 0.988 & 0.988 & 0.974 \\
    $L=800$  & 0.964 & 0.971 & 0.987 & 0.991 & 0.991 & 0.979 \\
    $L=1000$ & 0.970 & 0.975 & 0.992 & 0.996 & 0.995 & 0.989
\end{tabular}
\end{ruledtabular}
\end{table}

\newpage
\bibliographystyle{apsrev4-1}
\bibliography{refs}

\end{document}